\documentclass[preprint,12pt]{elsarticle}

\makeatletter
\def\ps@pprintTitle{%
    \let\@oddhead\@empty
    \let\@evenhead\@empty
    \def\@oddfoot{\footnotesize
         {\it{Submitted to Elsevier}} \hfill {\it{August 2025}}}%
    \let\@evenfoot\@oddfoot
    }
\makeatother

\usepackage{amssymb}
\usepackage[fleqn]{amsmath}
\usepackage{amsmath,amsthm,bm} 
\usepackage{amsfonts}
\usepackage{mathtools}
\usepackage{fdsymbol}
\usepackage{mathrsfs}
\usepackage{dutchcal} 

\usepackage{algorithm}
\usepackage{algpseudocode}

\usepackage[colorlinks=true, allcolors=blue]{hyperref}
\usepackage[margin=2.5cm,includefoot]{geometry} 
\usepackage{caption}
\usepackage{subcaption}
\usepackage{booktabs}  
\usepackage{longtable} 
\usepackage{multirow} 
\graphicspath{ {./Figures/} } 

\usepackage{soul,color}

\newcommand{\noop}[1]{}

\hypersetup{pdfauthor={Zihao Liu, Mohsen Ebrahimzadeh Hassanabadi, Daniel Dias-da-Costa}}

\journal{Journal of Sound and Vibration}

\usepackage{natbib}
\begin{document}
\begin{frontmatter}



\title{Experimental validation of universal filtering and smoothing for linear system identification using adaptive tuning}


\author[a]{Zihao Liu}
\author[b]{Sima Abolghasemi}
\author[a]{Mohsen Ebrahimzadeh Hassanabadi}
\author[b]{Nicholas E. Wierschem}
\author[a]{Daniel Dias-da-Costa\corref{cor1}}
\ead{daniel.diasdacosta@sydney.edu.au}
\cortext[cor1]{Corresponding author}

\address[a]{School of Civil Engineering, The University of Sydney, Sydney, NSW 2006, Australia}
\address[b]{Department of Civil and Environmental Engineering, University of Tennessee, Knoxville, TN 37996, USA}

\begin{abstract}

\noindent In Kalman filtering, unknown inputs are often estimated by augmenting the state vector, which introduces reliance on fictitious input models. In contrast, minimum-variance unbiased methods estimate inputs and states separately, avoiding fictitious models but requiring strict sensor configurations, such as full-rank feedforward matrices or without direct feedthrough. To address these limitations, two universal approaches have been proposed to handle systems with or without direct feedthrough, including cases of rank-deficient feedforward matrices. Numerical studies have shown their robustness and applicability, however, they have so far relied on offline tuning, and performance under physical sensor noise and structural uncertainties has not yet been experimentally validated. Contributing to this gap, this paper experimentally validates the universal methods on a five-storey shear frame subjected to shake table tests and multi-impact events. Both typical and rank-deficient conditions are considered. Furthermore, a self-tuning mechanism is introduced to replace impractical offline tuning and enable real-time adaptability. The findings of this paper provide strong evidence of the robustness and adaptability of the methods for structural health monitoring applications, particularly when sensor networks deviate from ideal configurations.
\end{abstract}

\end{frontmatter}

\section*{Highlights}
\begin{itemize}
\item Self-tuning is introduced as a replacement for offline tuning for real-time applications.
\item Universal filter and smoothing are experimentally validated on a shear frame.
\item Capability in rank-deficient systems is demonstrated using multi-impact tests.
\item The performance of the Universal methods is compared with state-of-the-art methods.
\item They show noise robustness and adaptability for structural health monitoring.
\end{itemize}

\begin{table}[ht]
    \centering
    \small
    \caption*{List of abbreviations}
    \label{tab1}
    \begin{tabular}{r l}
    \toprule
    AKF & Augmented Kalman filter \\
    DKF & Dual Kalman filter \\
    DF & Direct feedthrough \\
    GDF & Gillijns-and-De-Moor filter \\
    KF & Kalman filter \\
    MVU & Minimum-variance unbiased \\
    NDF & No direct feedthrough\\
    UF & Universal filter\\
    US & Universal smoothing \\
    NRMSE & Normalised root-mean-square error \\
    \bottomrule
    \end{tabular}
\end{table}

\section{Introduction} \label{introduction}

\noindent System identification plays a fundamental role in many engineering applications, including structural health monitoring~\cite{Bhowmik2020}, damage detection~\cite{Aral2018}, model updating~\cite{Moaveni2014}, and vibration control~\cite{Turco2017}. Among the various system identification techniques, vibration-based system identification has gained particular prominence due to its ability to extract dynamic characteristics from measured response data. These methods are typically classified into data-driven approaches, which leverage machine learning and deep learning techniques~\cite{Osama2017, Zhilu2021}, and model-based approaches, which incorporate physical models of the system to guide the estimation process~\cite{Jice2022, Raj2024}. Within the model-based family, Bayesian filtering methods, particularly the Kalman Filter (KF)~\cite{Kalman1960}, have become widely used for dynamic system identification. The KF has been applied to damage assessment in noisy environments~\cite{FengGao2006}, fatigue monitoring using sparse sensor networks~\cite{Papadimitriou2011}, and strain reconstruction in offshore structures subject to environmental loads~\cite{PengRen2014}. However, in these applications, the KF assumes that external inputs can be measured or can be treated as white noise, which limits its use in cases where the inputs are unknown or unmeasurable. To address this limitation, the Augmented Kalman Filter (AKF) was introduced~\cite{LourensReyndersEATAL2012}, allowing unknown inputs to be embedded directly into the state vector. This capability has enabled the AKF to be applied in a range of contexts, including the identification of impact forces in structural members~\cite{Saleem2019}, dynamic loading on vehicle suspension systems~\cite{RCumbo2019, HaoqiWang2019}, and unmeasured restoring forces in systems exhibiting nonlinear behaviour~\cite{Wang2021}. To address the so-called drift effect that arises when displacement-level measurements are unavailable, the Dual Kalman Filter (DKF) was proposed~\cite{EftekharAzam2015} and later validated experimentally~\cite{EftekharAzam2017}. Nevertheless, a critical limitation shared by both the AKF and DKF is their reliance on fictitious input evolution models, which introduce additional complexity in the tuning of process and measurement noise covariances, particularly when there are multiple inputs to be estimated.

To overcome the limitations associated with the fictitious input models used in augmented filtering approaches, \citet{Kitanidis1987}~explored an alternative method based on the principle of Minimum-Variance Unbiased (MVU) estimation. In this method, the system state can be estimated without requiring assumptions on the input dynamics, although the input itself is not directly estimated. This foundation was later extended to enable simultaneous estimation of both state and unknown input. \citet{Gillijns2007a} developed an MVU filter for systems without direct feedthrough in which acceleration-type measurements are not included in the estimation process. This estimator is widely referred to as the Gillijns-and-De-Moor Filter with No Direct Feedthrough (GDF-NDF). To allow for the inclusion of acceleration measurements, the \citet{Gillijns2007b} subsequently introduced a variant for systems with direct feedthrough, commonly known as the Gillijns-and-De-Moor Filter with Direct Feedthrough (GDF-DF). However, the GDF-DF requires that the feedforward matrix be full-rank, which means the number of acceleration measurements has to be equal to or greater than the number of unknown inputs. While both GDF-NDF and GDF-DF offer optimal estimation under ideal conditions, their implementation is limited by strict requirements on sensor configurations. In particular, the GDF-NDF cannot incorporate accelerometer data, which are among the most commonly used sensors in structural dynamics. Meanwhile, the GDF-DF becomes ill-conditioned or inapplicable when the feedforward matrix is rank-deficient -- a condition frequently encountered in practice where sensor networks are sparse. Although several improvements have been proposed to enhance numerical stability and estimation robustness~\cite{LourensPapadimitriouETAL2012, Martin2024}, the fundamental sensor type and placement constraints of GDF-based methods remain a significant barrier to their broader applicability in system identification.

To address the sensor configuration limitations of existing MVU estimators, the Universal Filter (UF)~\cite{Mohsen2023} was recently developed. Unlike previous methods, the UF can be applied to systems with or without direct feedthrough, including those with rank-deficient feedforward matrices. By reformulating the system inversion problem and utilising weighted least squares, the UF eliminates the need for restrictive assumptions on sensor placement while maintaining well-conditioned estimation across a wide range of system types.

A fundamental limitation shared by all filtering approaches, including the UF, is that they operate sequentially and rely solely on measurements available at a single timestep. This constraint makes filters sensitive to measurement noise and poor observability. As a result, system inversion can become ill-conditioned, adversely affecting estimation quality, especially in scenarios involving limited sensor coverage, displacement-only or strain-only measurements, or non-collocated sensor networks. To address this, smoothing techniques have been introduced to enhance estimation robustness by incorporating observations across a time window. In Kalman-type methods, smoothing has been explored through fixed-lag smoothing in AKF frameworks~\cite{Ulrika2019}, as well as window-based implementations of the extended KF~\cite{MingmingSong2023} and unscented KF~\cite{MingmingSong2022}. Smoothing algorithms have also been developed for MVU-based methods. \citet{Maes2018} proposed a GDF-based smoothing algorithm for systems with direct feedthrough, while \citet{Mohsen2022} addressed systems without direct feedthrough. In alignment with their filtering origins~\cite{Gillijns2007a, Gillijns2007b}, these MVU smoothing formulations remain subject to the same rank condition on the feedforward matrix, which limits their applicability to systems with full-rank configurations.

Building on the UF framework, the Universal Smoothing (US) method~\cite{Liu2025} was proposed to overcome these limitations. By incorporating measurement data from an extended observation window, the US improves estimation stability and accuracy while maintaining the same loose sensor network requirements. Although both the UF and US have shown promising results in numerical studies, their effectiveness under conditions involving sensor noise and model uncertainty has not yet been validated. In particular, their performance in the presence of physical sensor noise, imperfect observability, and model uncertainty remains to be consistently assessed.

Beyond the development of new estimation algorithms, substantial efforts have been given to improving the practical applicability of filtering-based methods through sensor placement strategies, tuning robustness, and real-time adaptability. For example, optimal sensor placement has been shown to improve KF performance~\cite{SongyeZhu2013}, while ill-conditioning in the AKF due to sensor layout has been addressed in~\cite{Tamarozzi2016}. Hyperparameter tuning and sensor deployment optimisation have also been investigated in~\cite{Aucejo2019} and~\cite{RCumbo2021}, respectively. To effectively utilise sensor data collected at different sampling rates, multi-rate filtering approaches have been introduced in several studies~\cite{Smyth2007, KiyoungKim2016, ZimoZhu2023}. In parallel, efforts to simplify filter tuning through direct noise covariance estimation have been explored~\cite{YuenKaVeng2013, StefanWernitz2022}, with the goal of supporting real-time applications. Recent advances include adaptive filters that dynamically select the number of retained modes in reduced-order models~\cite{XiaoHuaZhang2021}, and self-calibrating strategies that perform adaptive tuning either by estimating noise covariances directly~\cite{MingmingSong2020, Konstantinos2020} or by operating a bank of parallel filters~\cite{Vettori2023}.

Our study aims to experimentally validate the performance of the UF and US methods and investigate their practical benefits through an adaptive tuning scheme. The UF and US are evaluated through laboratory tests on a five-storey shear frame subject to both ground motion and multiple-impact excitations. A range of sensor configurations is considered, including arrangements with displacement-only and acceleration-only measurements, as well as rank-deficient and non-collocated arrangements that would typically violate the applicability conditions of conventional methods. In addition, a filter array–based adaptive tuning framework is implemented to enable real-time selection of the process noise covariance without requiring prior knowledge or offline calibration. The effectiveness of the proposed methods is assessed through comparisons with state-of-the-art methods, including the AKF and GDFs. To the best of the authors' knowledge, this is the first experimental study to validate the performance of UF and US under physical sensor noise, limited observability, and model uncertainty while also demonstrating their applicability to real-time structural health monitoring.

The remainder of this paper is organised as follows. Section~\ref{section2} introduces the theoretical formulation of the UF and US, including the adaptive tuning scheme based on a filter array. Section~\ref{section3} describes the experimental setup, test structure, instrumentation, and model identification procedures. Section~\ref{section4} presents the experimental results under various sensor configurations and loading conditions, along with a detailed evaluation of adaptive tuning performance and estimator convergence. Finally, section~\ref{section5} summarises the key findings, highlights the practical contributions of this work, and outlines directions for future research.

\section{Joint input and state estimation}
\label{section2}

\noindent In this section, the joint input-state estimation problem is formulated using a state-space representation of a linear structural system. The UF and US methods are then introduced, followed by the development of an adaptive tuning scheme based on a filter array approach. 

\subsection{Problem formulation and system equations} \label{section2_1}

\noindent The equation of motion for a structural system with $n_s$ degrees of freedom subject to $l$ applied loads can be expressed as:
\begin{equation} \label{eq:sys_eom}
    \mathbf{M}_s\ddot{\mathbf{u}}\left(t\right) + \mathbf{C}_s\dot{\mathbf{u}}\left(t\right) + \mathbf{K}_s\mathbf{u}\left(t\right) = \mathbf{S}\mathbf{p}(t),
\end{equation}

\noindent in which $\mathbf{M}_s\in\mathbb{R}^{n_s\times n_s}$, $\mathbf{C}_s\in\mathbb{R}^{n_s\times n_s}$ and $\mathbf{K}_s\in\mathbb{R}^{n_s\times n_s}$ are the mass matrix, damping matrix and stiffness matrix, respectively. The distribution of the dynamic loads $\mathbf{p}(t)\in\mathbb{R}^l$ is indicated by matrix $\mathbf{S}\in\mathbb{R}^{n_s\times l}$ and $\mathbf{u}\in\mathbb{R}^{n_s}$ is the displacement vector. The overdot denotes the derivative with respect to time, thus, $\dot{\mathbf{u}}\in\mathbb{R}^{n_s}$ and $\ddot{\mathbf{u}}\in\mathbb{R}^{n_s}$ are velocity and acceleration vectors, accordingly. If the system is subjected to ground acceleration, the term $\mathbf{Sp}(t)$ should be replaced by $-\mathbf{M}_s\mathbf{i}\ddot{\mathbf{u}}_g(t)$ where $\ddot{\mathbf{u}}_g(t)$ is the ground acceleration, and $\mathbf{i} \in \mathbb{R}^{n_s}$ is a unitary vector indicating the influence of ground acceleration to the structure. In the case modal properties or reduced order models are adopted, Eq.~\eqref{eq:sys_eom} can be rewritten by applying modal orthogonality considering an order of $n_r$ as: 
\begin{equation} \label{eq:rom_eom}
    \mathbf{M}_r\ddot{\mathbf{q}}(t)+\mathbf{C}_r\dot{\mathbf{q}}(t)+\mathbf{K}_r\mathbf{q}(t)=\mathbf{S}_r\mathbf{p}(t),
\end{equation}

\noindent and
\begin{equation} \label{eq:rom_matrices}
\begin{aligned}
    &\mathbf{M}_r=\mathbf{\Phi}^T\mathbf{M}_s\mathbf{\Phi},\\
    &\mathbf{C}_r=\mathbf{\Phi}^T\mathbf{C}_s\mathbf{\Phi},\\
    &\mathbf{K}_r=\mathbf{\Phi}^T\mathbf{K}_s\mathbf{\Phi},\\
    & \mathbf{B}_r=\mathbf{\Phi}^T\mathbf{S},
\end{aligned}
\end{equation}

\noindent in which, $\mathbf{q}\in\mathbb{R}^{n_r}$, $\dot{\mathbf{q}}\in\mathbb{R}^{n_r}$ and $\ddot{\mathbf{q}}\in\mathbb{R}^{n_r}$ are the dynamic responses in modal coordinates. $\mathbf{\Phi}\in\mathbb{R}^{n_s\times n_r}$ is the modal matrix containing eigenvectors of each mode. To utilise the data collected by sensors, Eq.~\eqref{eq:sys_eom} should be rewritten into a time-discrete form. Firstly, the equivalent ordinary differential equation of Eq.~\eqref{eq:sys_eom} can be obtained as:
\begin{equation} \label{eq:ode}
    \dot{\mathbf{x}}\left(t\right)=\mathbcal{A}\mathbf{x}\left(t\right)+\mathbcal{B}\mathbf{p}(t),
\end{equation}

\noindent where
\begin{equation} \label{eq:state}
    \mathbf{x}(t) = \begin{bmatrix} \mathbf{q}(t) \\ \dot{\mathbf{q}}(t)\\ \end{bmatrix},
\end{equation}
\vspace{-11pt}
\begin{equation} \label{eq:A_cal}
    \mathbcal{A} = \begin{bmatrix}
    \mathbf{0}_{n_r \times n_r} & \mathbf{I}_{n_r} \\
    -\mathbf{M}_{r}^{-1}\mathbf{K}_{r} & -\mathbf{M}_{r}^{-1}\mathbf{C}_{r} \\ \end{bmatrix},
\end{equation}
\vspace{-11pt}
\begin{equation} \label{eq:B_cal}
    \mathbcal{B} = \begin{bmatrix}
        \mathbf{0}_{n_r \times l} \\
        \mathbf{M}_{r}^{-1}\mathbf{S}_r \\ \end{bmatrix}.
\end{equation}

\noindent Note that $\mathbf{I}$ in the above equations is the identity matrix with the appropriate dimension. By using the matrix exponential approach~\cite{Brogan1991}, the time-discrete form of Eq.~\eqref{eq:ode} can be obtained:
\begin{equation} \label{eq:matrix_exp}
    \mathbf{x}_k=\mathbf{A}_{k-1}\mathbf{x}_{k-1}+\mathbf{B}_{k-1}\mathbf{p}_k,
\end{equation}

\noindent where, $\mathbf{x}_k = \mathbf{x}(t_k)$, $\mathbf{p}_k=\mathbf{p}(t_k)$, $\mathbf{A}=\exp \left(\mathbcal{A}\Delta t\right)$ and $\mathbf{B}=(\mathbf{A}-\mathbf{I}_{n_s})\mathbcal{A}^{-1}\mathbcal{B}$, with $t_k$ being $k$\textsuperscript{th} time instant and $\Delta t$ being the timestep size. 

The state-space model is established below based on the introduced structural system to implement recursive estimators. First, the process equation can be formulated as:
\begin{equation} \label{eq:process}
    \mathbf{x}_k=\mathbf{A}_{k-1}\mathbf{x}_{k-1}+\mathbf{B}_{k-1}\mathbf{p}_k+\mathbf{w}_{k-1},
\end{equation}

\noindent where the process noise $\mathbf{w}_{k}$ is assumed to be zero-mean white noise. The system matrices $\mathbf{A}_{k-1}$ and $\mathbf{B}_{k-1}$ are constant for a linear system. It should be highlighted that the zero-order hold assumption is made such that the input vector is expressed as $\mathbf{p}_k$. This is different from other recursive estimators, such as~\cite{Gillijns2007a, Gillijns2007b, Maes2018, Mohsen2022}, in which the input vector is $\mathbf{p}_{k-1}$, leading to one timestep lag. 

The measurement equation can be formulated next by defining an output vector $\mathbf{y}_k\in\mathbb{R}^q$ that collects sparse measurement of the system output, i.e., the dynamic responses of the structure:
\begin{equation} \label{eq:measurement}
    \mathbf{y}_k=\mathbf{C}_{k}\mathbf{x}_k+\mathbf{D}_{k}\mathbf{p}_k+\mathbf{v}_k,
\end{equation}

\noindent in which, $\mathbf{C}_{k}$ is the output matrix and $\mathbf{D}_{k}$ is the feedforward matrix. Considering linear systems and the measurement configuration remain unchanged during the estimation process, both matrices are constant. In addition, $\mathbf{v}_{k}$ is the measurement noise and is assumed to be zero-mean white noise. The output matrix $\mathbf{C}_{k}$ can be formulated as:
\begin{equation} \label{eq:output_matrix}
    \mathbf{C}_{k}=\mathfrak{C}
    \begin{bmatrix}
    \mathbf{I}_{n_r} & \mathbf{0}_{n_r\times n_r}\\
    \mathbf{0}_{n_r\times n_r} & \mathbf{I}_{n_r}\\
    -{\mathbf{M}_r}^{-1}\mathbf{K}_r&-{\mathbf{M}_r}^{-1}\mathbf{C}_r\\
    \end{bmatrix},
\end{equation}

\noindent where $\mathfrak{C}\in\mathbb{R}^{q\times 3n_r}$ is a Boolean matrix extracting the locations that are measured. Similarly, the feedforward matrix $\mathbf{D}_{k}$ is given by:
\begin{equation}\label{eq:feedforward_matrix}
    \mathbf{D}_{k}=\mathfrak{C}
    \begin{bmatrix}
    \mathbf{0}_{2n_r\times l}\\
    {\mathbf{M}_r}^{-1}\mathbf{S}_r\\
    \end{bmatrix}.
\end{equation}

Finally, the covariance matrices of process error and measurement noise read:
\begin{equation}
    \mathbf{Q}_{k} \triangleq \mathbb{E}\left[\mathbf{w}_{k}\mathbf{w}_{k}^T\right]
\end{equation}
\noindent and
\begin{equation}
    \mathbf{R}_{k} \triangleq \mathbb{E}\left[\mathbf{v}_{k}\mathbf{v}_{k}^T\right].
\end{equation}

\subsection{The Universal Filter method}

\noindent With the defined state-space model, the UF can be applied hereafter, which includes four steps, starting with the biased state estimation based on the information obtained from the previous timestep:
\begin{equation} \label{eq:uf_1}
    \hat{\mathbf{x}}_{k|k-1} = \mathbf{A}_{k} \hat{\mathbf{x}}_{k-1|k-1},
\end{equation}

\noindent where the subscript $k|k-1$ means the estimate at timestep $k$ based on the information up to timestep $k-1$. By comparing Eqs.~\eqref{eq:process} and~\eqref{eq:uf_1}, one can find that the bias is raised by the missing contribution of the input $\mathbf{p}_k$. Therefore, the innovation, i.e., the difference between the estimate $\hat{\mathbf{x}}_{k|k-1}$ and the measurement $\mathbf{y}_k$, implicitly defines the input. Therefore, the input can be estimated by using a gain matrix:
\begin{equation} \label{eq:uf_2}
    \hat{\mathbf{p}}_{k} = \mathbf{M}_{k}\left(\mathbf{y}_k-\mathbf{C}_{k}\hat{\mathbf{x}}_{k|k-1}\right).
\end{equation}

\noindent Note that the input gain matrix $\mathbf{M}_k$ is obtained by weighted least squares. Next, the a-priori state estimation is given by:
\begin{equation} \label{eq:uf_3}
    \hat{\mathbcal{x}}_{k|k-1} = \hat{\mathbf{x}}_{k|k-1} + \mathbf{B}_{k-1}\hat{\mathbf{p}}_{k}.
\end{equation}

Finally, the a-posteriori state estimation can be obtained with a state gain matrix $\mathbf{K}_k$ and measured data:
\begin{equation}\label{eq:uf_4}
    \hat{\mathbf{x}}_{k|k}=\hat{\mathbcal{x}}_{k|k-1}+\mathbf{K}_k\left(\mathbf{y}_k-\mathbf{C}_k\hat{\mathbcal{x}}_{k|k-1}-\mathbf{D}_k\hat{\mathbf{p}}_k\right).
\end{equation}

The filtering process, including error propagation, is summarised in Table~\ref{tab_uf} for implementation. Note that $(\cdot)^\dagger$ stands for the Moore–Penrose inverse (pseudoinverse).

\begin{table}[t]
    \centering
    \caption{Summary of the Universal Filter (UF)}
    \footnotesize
    \label{tab_uf}
    \begin{tabular}{l}
    \toprule
    \textbf{Initialisation}\\
    \hspace{3mm} • Define process and observation equations\\
    \hspace{3mm} • Assemble the error matrix: $\mathbcal{G}=\mathbf{CB}+\mathbf{D}$\\
    \hspace{3mm} • Assign initial values to $\hat{\mathbf{x}}_0$, $\mathbf{P}_0^{\mathbf{x}}$, $\mathbf{Q}_0$, $\mathbf{R}_0$ \\
    \midrule
    \textbf{Filtering loop}\\
    \hspace{3mm} • Input estimation\\
    \hspace{6mm} 1. $\hat{\mathbf{x}}_{k|k-1} = \mathbf{A}_{k-1} \hat{\mathbf{x}}_{k-1|k-1}$ \\
    \hspace{6mm} 2. $\mathbf{P}_k^{\mathbf{e}} = \mathbf{C}_k \left( \mathbf{A}_{k-1} \mathbf{P}_{k-1}^{\mathbf{x}} \mathbf{A}_{k-1}^T + \mathbf{Q}_{k-1} \right) \mathbf{C}_k^T + \mathbf{R}_k$ \\
    \hspace{6mm} 3. $\mathbf{P}_k^{\mathbf{p}}  = \left[\mathbcal{G}_k^T\left(\mathbf{P}_k^{\mathbf{e}}\right)^{\dagger}\mathbcal{G}_k\right]^{\dagger}$ \\
    \hspace{6mm} 4. $\mathbf{M}_k = \left[\mathbcal{G}_k^T\left(\mathbf{P}_k^{\mathbf{e}}\right)^{\dagger}\mathbcal{G}_k\right]^{\dagger}\mathbcal{G}_k^T\left(\mathbf{P}_k^{\mathbf{e}}\right)^{\dagger}$ \\
    \hspace{6mm} 5. $\hat{\mathbf{p}}_k = \mathbf{M}_k \left( \mathbf{y}_k-\mathbf{C}_k \hat{\mathbf{x}}_{k|k-1} \right)$ \\
    \hspace{3mm} • State estimation\\
    \hspace{6mm} 1. $\hat{\mathbcal{x}}_{k|k-1} = \hat{\mathbf{x}}_{k|k-1} + \mathbf{B}_{k-1}\hat{\mathbf{p}}_{k}$\\
    \hspace{6mm} 2. $\mathbf{P}_{k}^{\mathbf{xp}} = \left(\mathbf{P}_{k}^{\mathbf{px}}\right)^T -\mathbf{P}_{k-1}^{\mathbf{x}} \mathbf{A}_{k-1}^T \mathbf{C}_k^T \mathbf{M}_k^T$ \\
    \hspace{6mm} 3. $\mathbf{P}_k^{\mathbf{pw}} = \left( \mathbf{P}_k^{\mathbf{wp}} \right)^T = -\mathbf{M}_k\mathbf{C}_k\mathbf{Q}_{k-1}$ \\
    \hspace{6mm} 4. $\mathbf{P}_k^{\mathbcal{x}}  = \mathbf{A}_{k-1}\mathbf{P}_{k-1}^{\mathbf{x}}\mathbf{A}_{k-1}^T + \mathbf{B}_{k-1}\mathbf{P}_{k}^{\mathbf{p}}\mathbf{B}_{k-1}^T + \mathbf{Q}_{k-1}$ \\
    \hspace{24mm} $ + \mathbf{A}_{k-1}\mathbf{P}_{k}^{\mathbf{xp}}\mathbf{B}_{k-1}^T + \mathbf{B}_{k-1}\mathbf{P}_{k}^{\mathbf{px}}\mathbf{A}_{k-1}^T + \mathbf{B}_{k-1}\mathbf{P}_{k}^{\mathbf{pw}} + \mathbf{P}_{k}^{\mathbf{wp}}\mathbf{B}_{k-1}^T$ \\
    \hspace{6mm} 5. $\mathbf{P}_k^{\mathbf{p}\mathbcal{x}} = \mathbf{P}_k^{\mathbf{px}}\mathbf{A}_{k-1}^T + \mathbf{P}_k^{\mathbf{p}}\mathbf{B}_{k-1}^T + \mathbf{P}_k^{\mathbf{pw}}$ \\
    \hspace{6mm} 6. $\mathbf{P}_k^{\mathbf{pv}} = \left(\mathbf{P}_k^{\mathbf{vp}}\right)^T = -\mathbf{M}_k\mathbf{R}_k$ \\
    \hspace{6mm} 7. $\mathbf{P}_k^{\mathbf{v}\mathbcal{x}}  = \mathbf{P}_k^{\mathbf{vp}} \mathbf{B}_{k-1}^T$ \\
    \hspace{6mm} 8. $\mathbf{\Upsilon}_k  = \mathbf{C}_k\mathbf{P}_k^{\mathbcal{x}} + \mathbf{D}_k\mathbf{P}_k^{\mathbf{p}\mathbcal{x}} + \mathbf{P}_k^{\mathbf{v}\mathbcal{x}}$ \\
    \hspace{6mm} 9. $\mathbf{\Psi}_k =\mathbf{C}_k\mathbf{\Upsilon}_k^T + \mathbf{D}_k\left(\mathbf{P}_k^{\mathbf{p}\mathbcal{x}}\mathbf{C}_k^T+\mathbf{P}_k^{\mathbf{p}}\mathbf{D}_k^T+\mathbf{P}_k^{\mathbf{pv}} \right)+\mathbf{P}_k^{\mathbf{v}\mathbcal{x}}\mathbf{C}_k^T + \mathbf{P}_k^{\mathbf{vp}}\mathbf{D}_k^T +\mathbf{R}_k$ \\
    \hspace{6mm} 10. $\mathbf{K}_k = \mathbf{\Upsilon}_k^T\mathbf{U}_k\left(\mathbf{U}_k^T\mathbf{\Psi}_k\mathbf{U}_k\right)^{-1}\mathbf{U}_k^T$ \\
    \hspace{6mm} 11. $\mathbf{P}_{k}^{\mathbf{x}} = \mathbf{K}_k\mathbf{\Psi}_k\mathbf{K}_k^T - \mathbf{\Upsilon}_k^T\mathbf{K}_k^T - \mathbf{K}_k\mathbf{\Upsilon}_k + \mathbf{P}_k^{\mathbcal{x}}$ \\
    \hspace{6mm} 12. $\hat{\mathbf{x}}_{k|k} = \hat{\mathbcal{x}}_{k|k-1} + \mathbf{K}_k \left( \mathbf{y}_k - \mathbf{C}_k\hat{\mathbcal{x}}_{k|k-1}-\mathbf{D}_k\hat{\mathbf{p}}_{k} \right)$ \\
    \bottomrule
    \end{tabular}
\end{table}

\subsection{The Universal Smoothing method}
\noindent One noticeable feature of the filtering method is that each filtering loop can only utilise the observation data at one single timestep. This limitation can lead to ill-conditioned estimation results when the measurements contain a high level of noise, the number of sensors is insufficient, or the combination and placement of sensors are not ideal. This ill-conditionedness of filtering methods can be lifted by smoothing techniques. Unlike filters, smoothing methods can include measurements obtained in a time window into the estimation results, thereby resulting in improved estimation quality. The process of the US method is briefly introduced hereafter.

Since smoothing methods can access measurements collected in a time window, the measurement equation,  Eq.~\eqref{eq:measurement}, must be extended by introducing a time window size $N$:
\begin{equation}\label{eq:extended_measurement}
\mathbcal{y}_k=\mathbcal{C}_k\mathbf{x}_k+\mathbcal{D}_k\mathbcal{p}_k+\mathbcal{H}_k\mathbcal{w}_k+\mathbcal{v}_k,
\end{equation}

\noindent where $\mathbcal{y}_k\in\mathbb{R}^{(N+1)q}$ is the extended measurement vector that contains measured data from timestep $k$ up to timestep $k+N$. Similarly, $\mathbcal{p}_k$, $\mathbcal{w}_k$ $\mathbcal{v}_k$ are the extended input, process error and measurement noise vectors, respectively. The construction of these extended state-space matrices involved in the smoothing process can be found in~\ref{AppendixA}.

The US also starts with a biased state estimation, i.e. Eq.~\eqref{eq:uf_1}. However, with the extended measurement, the innovation becomes $\mathbcal{y}_k-\mathbcal{C}_k\hat{\mathbf{x}}_{k|k-1}$. Therefore, the input estimation reads as:
\begin{equation}\label{eq:us_2}\hat{\mathbcal{p}}_k=\mathbf{M}_k\left(\mathbcal{y}_k-\mathbcal{C}_k\hat{\mathbf{x}}_{k|k-1}\right).
\end{equation}

Note that the extended input vector $\hat{\mathbcal{p}}_k$ contains inputs in the time window, in which case the input at the timestep $k$ can be extracted by:
\begin{equation}
\hat{\mathbf{p}}_k=\begin{bmatrix}\mathbf{I}_m & \bm{\mathbf{0}}_{m\times Nm}\\ \end{bmatrix}
\hat{\mathbcal{p}}_k.
\end{equation}

Next, the a-priori state estimation follows Eq.~\eqref{eq:uf_3} and with the extended measurement equation, the final a-posteriori estimation can be obtained by:
\begin{equation}
    \hat{\mathbf{x}}_{k|k}=\hat{\mathbcal{x}}_{k|k-1}+\mathbf{K}_k\left(\mathbcal{y}_k-\mathbcal{C}_k\hat{\mathbcal{x}}_{k|k-1}-\mathbcal{D}_k\hat{\mathbcal{p}}_k\right).
\end{equation}

The detailed smoothing process is summarised in Table~\ref{tab_us}. 
\clearpage
{
\footnotesize
\begin{longtable}[t]{l}
    \caption{Summary of the Universal Smoothing (US)}
    \label{tab_us}\\

    \toprule
    \multicolumn{1}{l}{\textbf{Initialisation}}\\
    \endfirsthead
    \midrule
    \endfoot
    
    \multicolumn{1}{c}{Continuation of Table \ref{tab_us}}\\
    \midrule
    \endhead
    \bottomrule
    \endlastfoot

    \hspace{3mm} • Define process, observation and extended observation equations\\
    \hspace{3mm} • Assign initial values to $\hat{\mathbf{x}}_{0|0}$, $\mathbf{P}_0$, $\mathbf{P}_0^{\mathbf{xw}}$, $\mathbf{P}_0^{\mathbf{xv}}$, $\mathbf{Q}_0$, $\mathbf{R}_0$ and build $\mathbcal{Q}_{0,0}$, $\mathbcal{Q}_{0,1}$, $\mathbcal{R}_{0,0}$, $\mathbcal{R}_{0,1}$\\
    \hspace{3mm} • Build time-independent matrices $\breve{\mathbf{\bm{\varepsilon}}}_i\triangleq
                     \begin{bmatrix}\mathbf{0}_{iN\times i} & \mathbf{I}_{Ni}\\
                     \mathbf{0}_{i\times i} & \mathbf{0}_{i\times i N}\\\end{bmatrix}$, $\bar{\mathbf{\bm{\varepsilon}}}_i\triangleq
                     \begin{bmatrix}\mathbf{0}_{iN\times i N} & \mathbf{0}_{iN\times i}\\
                     \mathbf{0}_{i\times i N} & \mathbf{I}_i\\\end{bmatrix}$\\
    \midrule
    \textbf{Smoothing loop}\\
    \hspace{3mm} • Build matrices for input estimation\\
    \hspace{6mm} 1. $\breve{\mathbf{D}}_k\triangleq\mathbcal{D}_k+\mathbcal{C}_k\mathbf{B}_{k-1}\begin{bmatrix}\mathbf{I}_l & \bm{\mathbf{0}}_{l\times N l}\\ \end{bmatrix}$\\
    \hspace{6mm} 2. $\breve{\mathbf{H}}_k\triangleq\mathbcal{H}_k+\begin{bmatrix}\mathbcal{C}_k & \bm{\mathbf{0}}_{\left(N+1\right)q\times N 2n_r}\\ \end{bmatrix}$\\
    \hspace{6mm} 3. $\mathbf{\bm{\Sigma}}_k\triangleq
                     \begin{bmatrix}\mathbcal{C}_k\mathbf{A}_{k-1} & \breve{\mathbf{H}}_k & \mathbf{I}_{(N+1)q}\\ \end{bmatrix}$ \\
    \hspace{3mm} • Obtain input estimate\\
    \hspace{6mm} 1. $\hat{\mathbf{x}}_{k|k-1}=\mathbf{A}_{k-1}\hat{\mathbf{x}}_{k-1|k-1}$\\
    
    \hspace{6mm} 2. $\mathbf{P}_k^\mathbcal{p}=\left[\breve{\mathbf{D}}_k^T\left(\mathbf{\bm{\Sigma}}_k\mathbf{\bm{\Lambda}}_k\mathbf{\bm{\Sigma}}_k^T\right)^{-1}\breve{\mathbf{D}}_k\right]^{-1}$\\
    \hspace{6mm} 3. $\mathbf{P}_k^\mathbf{p}=\begin{bmatrix}\mathbf{I}_l & \bm{\mathbf{0}}_{l\times Nl}\\ \end{bmatrix}\mathbf{P}_k^\mathbcal{p}\begin{bmatrix}\mathbf{I}_l & \bm{\mathbf{0}}_{l\times Nl}\\ \end{bmatrix}^T$\\
    \hspace{6mm} 4. $\mathbf{M}_k=\mathbf{P}_k^\mathbcal{p}\breve{\mathbf{D}}_k^T\left(\mathbf{\bm{\Sigma}}_k\mathbf{\bm{\Lambda}}_k\mathbf{\bm{\Sigma}}_k^T\right)^{-1}$\\
    \hspace{6mm} 5. $\hat{\mathbcal{p}}_k=\mathbf{M}_k\left(\mathbcal{y}_k-\mathbcal{C}_k\hat{\mathbf{x}}_{k|k-1}\right)$\\
    \hspace{6mm} 6. $\hat{\mathbf{p}}_k=\begin{bmatrix}\mathbf{I}_l & \bm{\mathbf{0}}_{m\times Nl}\\ \end{bmatrix}\hat{\mathbcal{p}}_k$\\
    \hspace{3mm} • Build matrices for state estimation\\
    \hspace{6mm} 1. $\mathbf{V}_k\triangleq\mathbf{B}_{k-1}\begin{bmatrix}\mathbf{I}_l & \bm{\mathbf{0}}_{l\times Nl}\\ \end{bmatrix}\mathbf{M}_k$\\
    \hspace{6mm} 2. $\mathbf{W}_k\triangleq-\mathbf{V}_k\breve{\mathbf{H}}_k+\begin{bmatrix}\mathbf{I}_{2n_r} & \bm{\mathbf{0}}_{2n_r\times N2n_r}\\ \end{bmatrix}$\\
    \hspace{6mm} 3. $\breve{\mathbf{A}}_{k-1}\triangleq\mathbf{A}_{k-1}-\mathbf{V}_k\mathbcal{C}_k\mathbf{A}_{k-1}$\\
    \hspace{6mm} 4. $\mathbf{\bm{\Pi}}_k\triangleq\begin{bmatrix}\breve{\mathbf{A}}_{k-1} & \mathbf{W}_k & -\mathbf{V}_k\\ \end{bmatrix}$\\
    \hspace{6mm} 5. $\mathbf{\bm{\Theta}}_k\triangleq\mathbcal{D}_k\mathbf{M}_k$\\
    \hspace{6mm} 6. $\mathbf{\Omega}_k\triangleq
    \begin{bmatrix} \mathbcal{C}_k\breve{\mathbf{A}}_{k-1}-\mathbf{\bm{\Theta}}_k\mathbcal{C}_k\mathbf{A}_{k-1}
    &\mathbcal{C}_k\mathbf{W}_k-\mathbf{\bm{\Theta}}_k\breve{\mathbf{H}}_k+\mathbcal{H}_k
    &-\mathbcal{C}_k\mathbf{V}_k-\mathbf{\bm{\Theta}}_k+\mathbf{I}_{\left(N+1\right)q\times 2n_r}\\ \end{bmatrix}$\\
    \hspace{3mm} • Obtain state estimate\\
    \hspace{6mm} 1. $\mathbf{P}_k^\mathbcal{x}=\mathbf{\bm{\Pi}}_k\mathbf{\bm{\Lambda}}_k\mathbf{\bm{\Pi}}_k^T$\\
    \hspace{6mm} 2. $\mathbf{\bm{\Upsilon}}_k=-\mathbf{\bm{\Omega}}_k\mathbf{\bm{\Lambda}}_k\mathbf{\bm{\Pi}}_k^T$\\
    \hspace{6mm} 3. $\mathbf{\bm{\Phi}}_k=\mathbf{\bm{\Omega}}_k\mathbf{\bm{\Lambda}}_k\mathbf{\bm{\Omega}}_k^T$\\
    \hspace{6mm} 4. $\mathbf{K}_k=-\mathbf{\bm{\Upsilon}}_k^T\mathbf{U}_k^T\left(\mathbf{U}_k\mathbf{\bm{\Phi}}_k\mathbf{U}_k^T\right)^{-1}\mathbf{U}_k$\\
    \hspace{6mm} 5. $\hat{\mathbcal{x}}_{k|k-1}=\hat{\mathbf{x}}_{k|k-1}+\mathbf{B}_{k-1}\hat{\mathbf{p}}_k$\\
    \hspace{6mm} 6. $\hat{\mathbf{x}}_{k|k}=\hat{\mathbcal{x}}_{k|k-1}+\mathbf{K}_k\left(\mathbcal{y}_k-\mathbcal{C}_k\hat{\mathbcal{x}}_{k|k-1}-\mathbcal{D}_k\hat{\mathbcal{p}}_k\right)$\\
    \hspace{6mm} 7. $\mathbf{P}_k=\mathbf{P}_k^\mathbcal{x}+\mathbf{K}_k\mathbf{\bm{\Upsilon}}_k+\mathbf{\bm{\Upsilon}}_k^T\mathbf{K}_k^T+\mathbf{K}_k\mathbf{\bm{\Phi}}_k\mathbf{K}_k^T$\\
    \hspace{3mm} • Update covariance matrices linked to input estimation for the next loop\\
    \hspace{6mm} 1. $\mathbf{T}_k=\mathbf{I}_{\left(N+1\right)d\times n}-\mathbf{K}_k\mathbcal{C}_k$\\
    \hspace{6mm} 2. $\mathbcal{W}_k = \mathbf{T}_k\mathbf{W}_k+\mathbf{K}_k\mathbf{\bm{\Theta}}_k\breve{\mathbf{H}}_k-\mathbf{K}_k\mathbcal{H}_k$\\
    \hspace{6mm} 3. $\mathbcal{V}_k=-\mathbf{T}_k\mathbf{V}_k+\mathbf{K}_k\mathbf{\bm{\Theta}}_k-\mathbf{K}_k$\\
    \hspace{6mm} 4. $\tilde{\mathbcal{A}}_k = \mathbf{T}_k\breve{\mathbf{A}}_{k-1}+\mathbf{K}_k\mathbf{\bm{\Theta}}_k\mathbcal{C}_k\mathbf{A}_{k-1}$ \\
    \hspace{6mm} 5. $\mathbf{P}_k^{\mathbf{xw}}=\tilde{\mathbcal{A}}_k\mathbf{P}_{k-1}^{\mathbf{xw}}\breve{\mathbf{\bm{\varepsilon}}}_n^T+\mathbcal{W}_k\mathbcal{Q}_{k,k+1}$\\
    \hspace{6mm} 6. $\mathbf{P}_k^{\mathbf{xv}}=\tilde{\mathbcal{A}}_k\mathbf{P}_{k-1}^{\mathbf{xv}}\breve{\mathbf{\bm{\varepsilon}}}_d^T+\mathbcal{V}_k\mathbcal{R}_{k,k+1}$\\
 \end{longtable}
}

\subsection{Adaptive filtering scheme} \label{section2_4}

\noindent Previous studies carried out the tuning of the process and measurement noise covariance matrices, $\mathbf{Q}_k$ and $\mathbf{R}_k$, offline~\cite{Mohsen2023, Liu2025}. This approach is commonly adopted in many studies~\cite{LourensReyndersEATAL2012, EftekharAzam2015}, where candidate values for $\mathbf{Q}_k$ are selected using grid search techniques and regularisation methods, such as the L-curve, while the measurement noise covariance, $\mathbf{R}_k$, was fixed to the true root-mean-square error of the measurements, which is typically known in the numerical simulated experiments. Although these offline tuning techniques can yield satisfactory performance under nominal conditions, they require manual intervention and do not account for time-varying uncertainties or modelling discrepancies that arise in practical applications. This limitation highlights the need for adaptive tuning strategies that can automatically adjust the noise covariance matrices in real-time in response to changing operating conditions. 

\begin{figure}[ht]
    \centering
    \includegraphics[trim={0cm 3.5cm 0cm 3cm}, clip, width=0.99\textwidth]{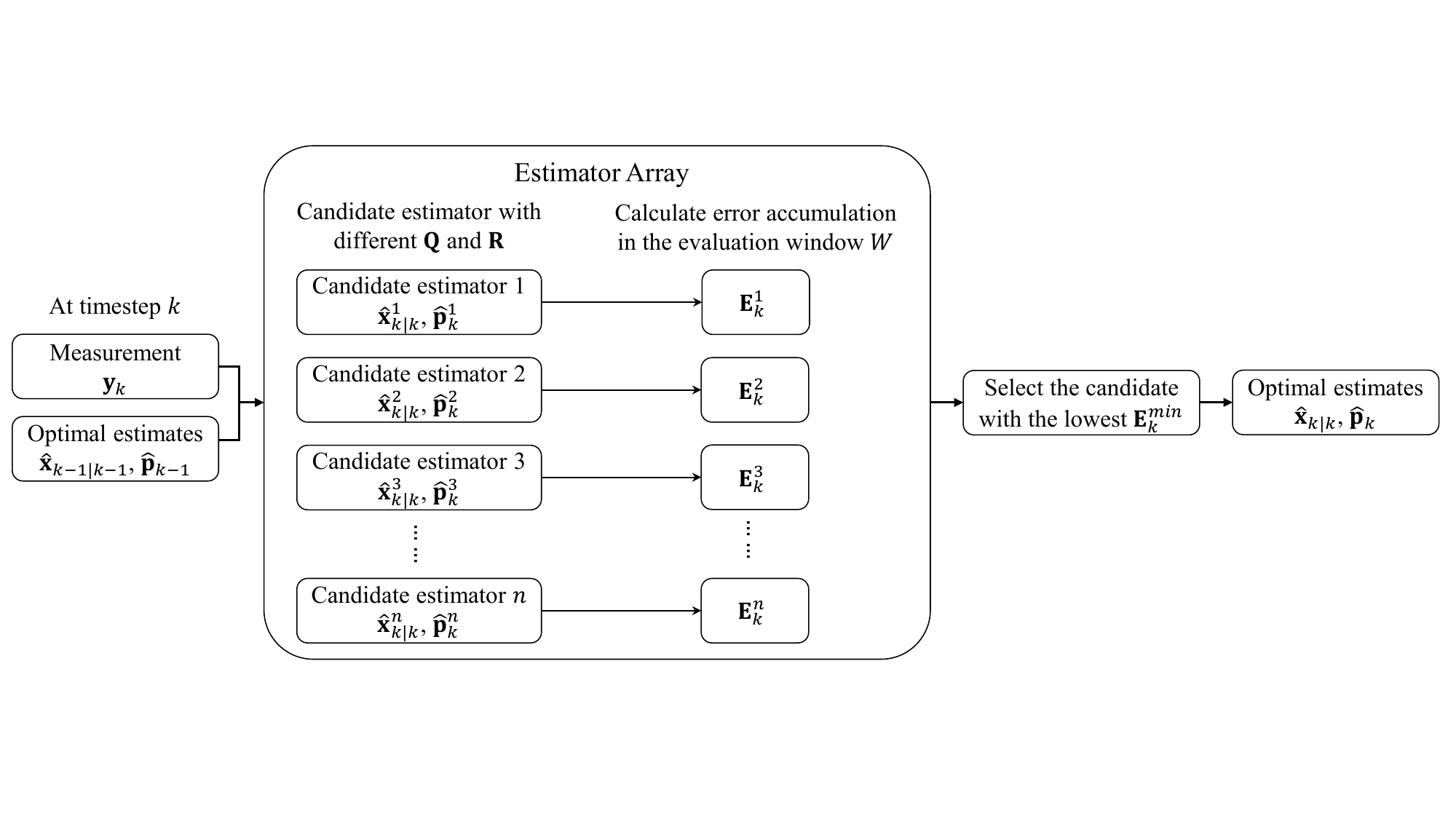}
    \caption{Adaptive tuning through an estimator array.}
    \label{fig:array}
\end{figure}

Motivated by this need, an adaptive tuning scheme is proposed in this paper, which is achieved by deploying a filter array in which each candidate filter is configured with distinct $\mathbf{Q}_k$ and $\mathbf{R}_k$ matrices sampled over a logarithmically uniform range. These candidates represent various hypotheses regarding the underlying process noise characteristics. At each time step, all candidate filters update their input and state estimates in parallel, and an error metric is computed for each filter. The error metric is derived from the innovation and accumulated over a predefined window:
\begin{equation}
    \mathbf{E}_{k} = \sum_{i=1}^{W}
    \bigl\lVert 
      \mathbf{y}_{k-W+i}-\mathbf{C}_k\hat{\mathbcal{x}}_{k-W+i|k-W+i-1}-\mathbf{D}_k\hat{\mathbf{p}}_{k-W+i}
    \bigr\rVert^2,
\end{equation}

\noindent where $W$ is the length of the evaluation window. At the end of each timestep, the candidate filter with the lowest cumulative error is selected as the optimal model, and its state and covariance estimates are used to reinitialise the entire filter array for the next cycle. This multi-model framework not only mitigates the risk of numerical instability inherent in a simultaneous continuous adaptation of $\mathbf{Q}_k$ and $\mathbf{R}_k$, but also enhances the ability of the filter to track changes in process noise under uncertain or nonstationary conditions. The concept of the filter array is illustrated in Fig.~\ref{fig:array}.

In the following validation studies, the adaptive tuning scheme was applied with variation only in the process noise covariance matrix $\mathbf{Q}_k$. The measurement noise covariance matrix $\mathbf{R}_k$ was fixed throughout, with its diagonal values set based on the known signal-to-noise ratio of the sensors. This approach reflects practical scenarios where sensor noise characteristics are known a priori, either from calibration or manufacturer specifications. Fixing one of the error covariance matrices also improves numerical stability during online tuning and is sufficient for accurate state estimation when measurement noise is well-characterised~\cite{MingmingSong2020}.

It is noted that alternative approaches, such as direct covariance estimation methods based on innovation statistics, have been widely investigated for adaptive tuning of noise covariances. However, such methods typically rely on the assumption of stationary noise environments and require careful tuning of additional parameters to maintain positive definiteness. In nonstationary or impulsive environments, these direct estimation techniques are prone to divergence or numerical instability~\cite{XiangxiangDong2022, ZengkeLi2025}. By contrast, the filter array approach adopted in this study circumvents these issues by discretising possible noise hypotheses and allowing selection based on observed data, thereby providing a more robust and transparent adaptation mechanism. This design choice ensures stable performance under changing operating conditions without introducing complex additional tuning parameters.

\section{Experimental validation} \label{section3}
\noindent In this section, the experiment undertaken to validate the UF and US is presented. A five-storey shear frame structure was tested under two distinct excitation scenarios: ground motion using a shake table and multiple-impact loading using an instrumented hammer. The objective was to evaluate the performance and robustness of the UF and US under various operating conditions, including varying sensor configurations and rank-deficient feedforward matrices. The details of the test structure, instrumentation layout, excitation plans, and determination of the modal properties of the test structure are described hereafter.

\subsection{Test structure and instrumentations}
\begin{figure}[ht]
    \centering
    \begin{subfigure}{0.45\textwidth}
        \caption{} \label{fig:structure}
        \includegraphics[trim={0cm 0cm 0cm 0cm}, clip, width=\textwidth]{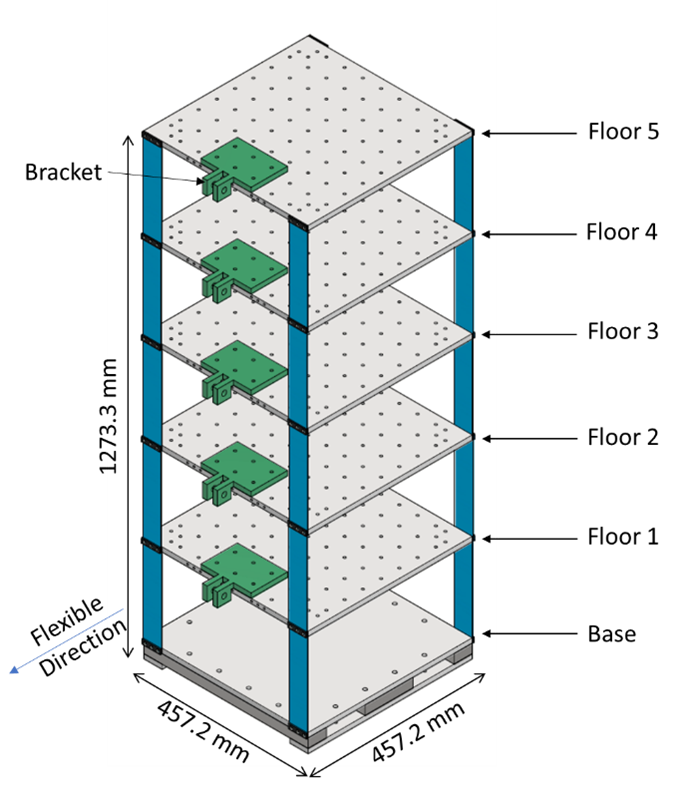}
    \end{subfigure}
    \begin{subfigure}{0.36\textwidth}
        \caption{} \label{fig:instrumentations}
        \includegraphics[trim={0cm 0cm 0cm 0cm}, clip, width=\textwidth]{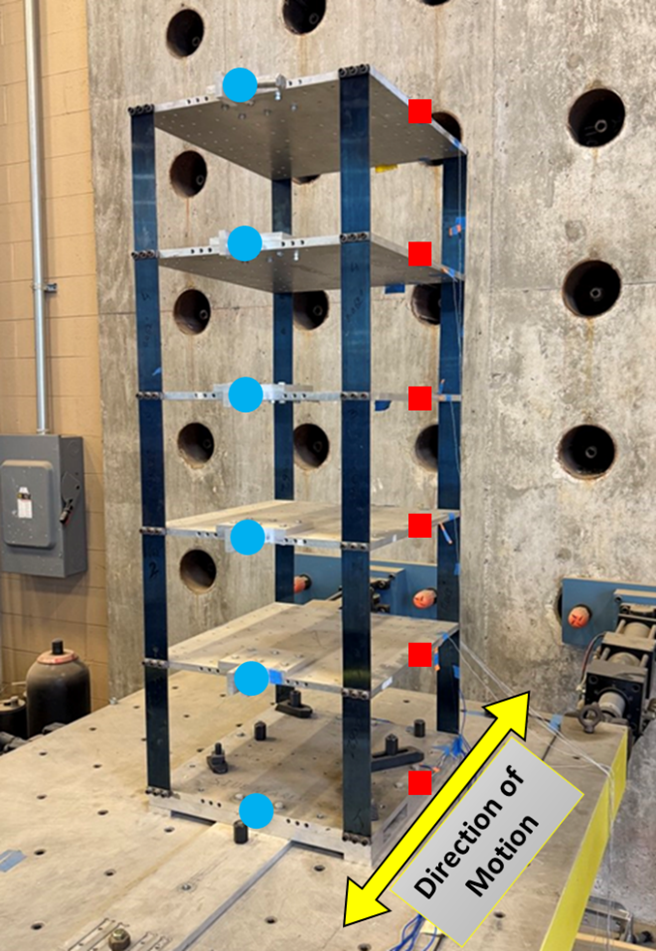}
    \end{subfigure}
    \label{fig:test_structure}
    \caption{(a) 3D design model of the test structure, (b) the five-story shear frame on the shake table where the red squares indicate the placement of Optotrak markers and the blue circles represent the side opposite to the placement of accelerometers.}
\end{figure}

\noindent The physical model is shown in Fig.~\ref{fig:structure}, which consists of a base and floor plates made of aluminium, each measuring $457.2\ \rm{mm} \times 457.2\ \rm{mm} \times 12.7\ \rm{mm}$. The mass of each floor, including the brackets that were attached to each floor plate of the structure for a separate set of tests, is $8.083$ kg. The columns, positioned at each corner and providing flexibility primarily in one direction, are made of grade 1095 spring steel with a width of $50.8$~mm, a thickness of $1.575$ mm, and an individual mass of $0.154$~kg. The flexural rigidity of each column about its weak axis, determined through a three-point bending test, is $\rm{EI}=3.205\ \rm{Nm^2}$, which corresponds to a floor stiffness of $1.24\times10^4\ \rm{N/m}$. The centre-to-centre height of each story is 244.5~mm, while the clear story height is $231.8$~mm. Further details of this five-story structure are provided in the design drawings of the structure~\cite{Lifsey2023}.

The experimental setup included instrumentation for measuring both acceleration and displacement to capture the dynamic response of the structure. Fig.~\ref{fig:instrumentations} illustrates the placement of sensors. Six PCB model 352C33SN accelerometers were installed at the base and each floor plate, positioned along the flexible direction and oriented to measure in the direction of loading to record the acceleration of both the ground (shake table) and the floors. Accelerometer and instrumented hammer data were acquired with a 2560 Hz sampling rate. 
Additionally, six Optotrak markers were installed on the opposite face of the structure to measure the absolute position of the base and each floor. These markers are part of the optical tracking system from NDI, which detects emitted infrared light for precise motion tracking and provides three-dimensional position records. These sensors were placed at the midpoint of the side of each floor plate to ensure accurate tracking of the motion and utilised a sampling rate of 100 Hz with a separate data acquisition system. 
To achieve synchronisation between the two data acquisition systems, the system linked to the acceleration sensors served as a global reference timer. Each time the Optotrak system captured a displacement measurement, it sent a voltage signal to the data acquisition system collecting acceleration measurements, marking the corresponding time for each displacement record. The timestamps for the displacement and acceleration measurements were then aligned. The acceleration data was down-sampled to 100 Hz to match the sampling rate of the displacement measurements. 

\subsection{Test setup}
\begin{figure}[ht]
    \centering
    \begin{subfigure}{0.45\textwidth}
        \caption{} \label{fig:gournd_acc}
        \includegraphics[trim={3cm 10.5cm 4cm 11.5cm}, clip, width=\textwidth]{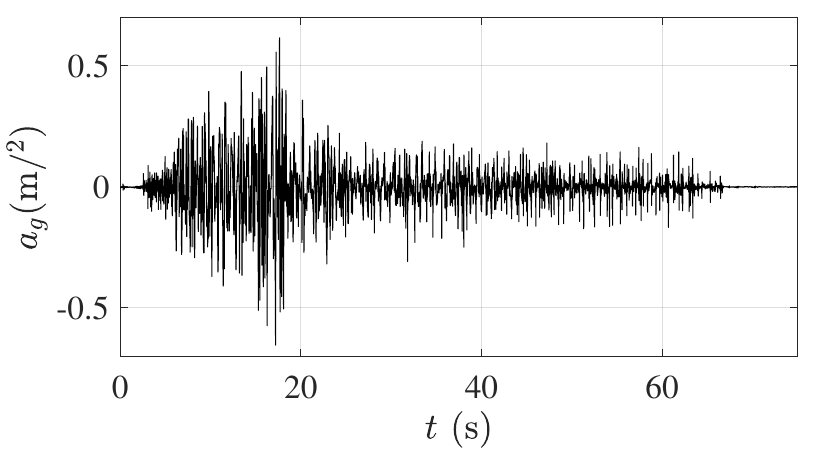}
    \end{subfigure}
    \begin{subfigure}{0.45\textwidth}
        \caption{} \label{fig:impacts}
        \includegraphics[trim={3cm 10.5cm 4cm 11.5cm}, clip, width=\textwidth]{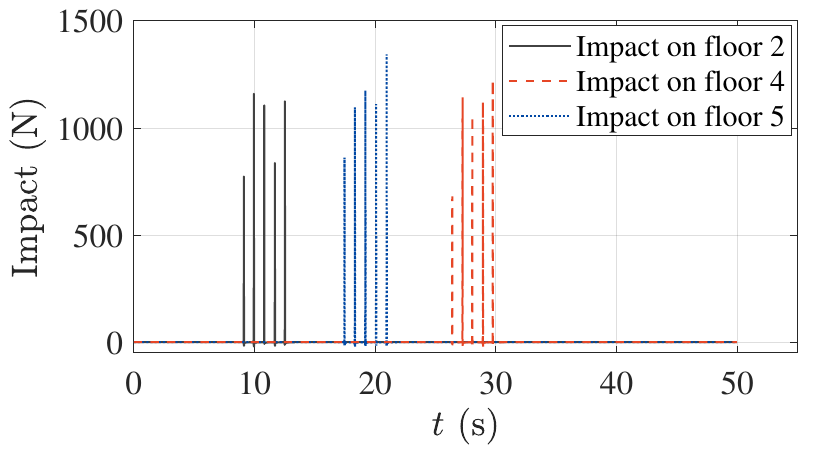}
    \end{subfigure}
    \label{fig:experiments}
    \caption{Time histories of (a) ground motion performed by the shake table and (b) impact loads applied on different floors.}
\end{figure}

\noindent The five-story structure was bolted to a $1.2\ \rm{m} \times 1.2\ \rm{m}$ six-degree-of-freedom shake table at the University of Tennessee. For validation purposes, two types of excitation were considered: ground excitation and multiple-floor excitation. In both cases, the excitation was applied in the flexible direction of the structure.
The 1999 Chi-Chi earthquake in Taiwan, specifically the first horizontal component of the ground acceleration measured by station CHY101, was selected as the ground excitation for this experimental work. This record was obtained from the Pacific Earthquake Engineering Research Center (PEER) ground motion database~\cite{Timothy2014}. The ground motion was applied as a unidirectional horizontal motion by the shake table that was aligned with the flexible direction of the structure. The recorded motion was scaled down to 40\%, ensuring significant structural response without causing damage to the physical model. While the shake table could not perfectly replicate the ground motion, an iterative process was employed prior to testing to refine the shake table commands, effectively reproducing the desired ground motion. The time history of the actual performed ground acceleration by the shake table is presented in Fig.~\ref{fig:gournd_acc}.

The structure was also subjected to an input consisting of multiple hammer strikes using a PCB model 086C03 instrumented hammer at various floors of the structure without ground excitation applied. Different sequences of instrumented hammer strikes were applied to implement multiple excitations. In one loading pattern, excitations were applied sequentially by tapping five times at Floor 2, then Floor 5, and finally Floor 4, with a one-second interval between each tap. The time history of the performed impact loads is shown in Fig.~\ref{fig:impacts}.

\subsection{State-space model identification}

\noindent As discussed in Section~\ref{section2_1}, a state-space model of the monitored structure must be obtained to implement recursive input-state estimators. In this paper, two methods are used to identify the state-space model and also extract modal properties of the test structure. The first method is called the Transformed Stochastic Subspace Identification method (T-SSID)~\cite{ChatzisTSSID2015}, and the subspace state-space model is then transformed back to the physical coordinates. The modal properties are obtained by eigenanalysis. The T-SSID approach first employs a subspace identification algorithm, such as N4SID, to derive an initial state-space model from measured acceleration, displacement, and ground motion data. This model, initially expressed in abstract coordinates, is then transformed into physical coordinates through a carefully determined transformation matrix, ensuring that the resulting system matrices accurately represent the underlying structural dynamics. Once in physical form, eigenvalue analysis is applied to the state matrix to extract the modal properties of the structure. The eigenvalues yield the natural frequencies and damping ratios -- calculated from the magnitude and real parts of the continuous-time eigenvalues -- while the associated eigenvectors provide the mode shapes, which can be normalised as needed for further interpretation or comparison with theoretical predictions.

The second method begins by computing frequency response functions (FRFs) from measured dynamic response data, such as acceleration and displacement, using spectral analysis techniques. Modal properties are then extracted from the FRF by identifying resonance peaks and applying the half-power bandwidth method~\cite{Brownjohn2003} to determine the natural frequencies and damping ratios, while the mode shapes are inferred from the spatial distribution of the response amplitudes. The data collected from the shake table test is used to obtain modal properties, and Table~\ref{tab-config-exp} compares the modal properties identified by the T-SSID and FRF approaches, and the differences are shown to be minimal. In the following case studies, the modal properties obtained by the T-SSID method are used to construct the state-space model of the structure.
\begin{table}[ht]
    \centering
    \caption{Modal properties of the test structure.}
    \label{tab-config-exp}
    \footnotesize{
    \begin{tabular}{c c c c c c c}
    \toprule
    \multicolumn{2}{c}{Modal properties} & Mode 1 & Mode 2 & Mode 3 & Mode 4 & Mode 5 \\
    \midrule
    \multirow{2}{12em}{Natural frequencies (Hz)} & T-SSID & 1.573 & 4.789 & 7.625 & 9.834 & 10.619 \\
    & FRF & 1.574 & 4.789 & 7.631 & 9.832 & 11.286\\
    \midrule
    \multirow{2}{12em}{Modal damping coefficients} & T-SSID & 0.016 & 0.014 & 0.016 & 0.017 & 0.016 \\
    & FRF & 0.016 & 0.017 & 0.012 & 0.012 & 0.014\\
    \bottomrule
    \end{tabular}}
\end{table}

\begin{figure}[ht]
    \centering
    \includegraphics[trim={3cm 9cm 4cm 9cm}, clip, width=0.5\textwidth]{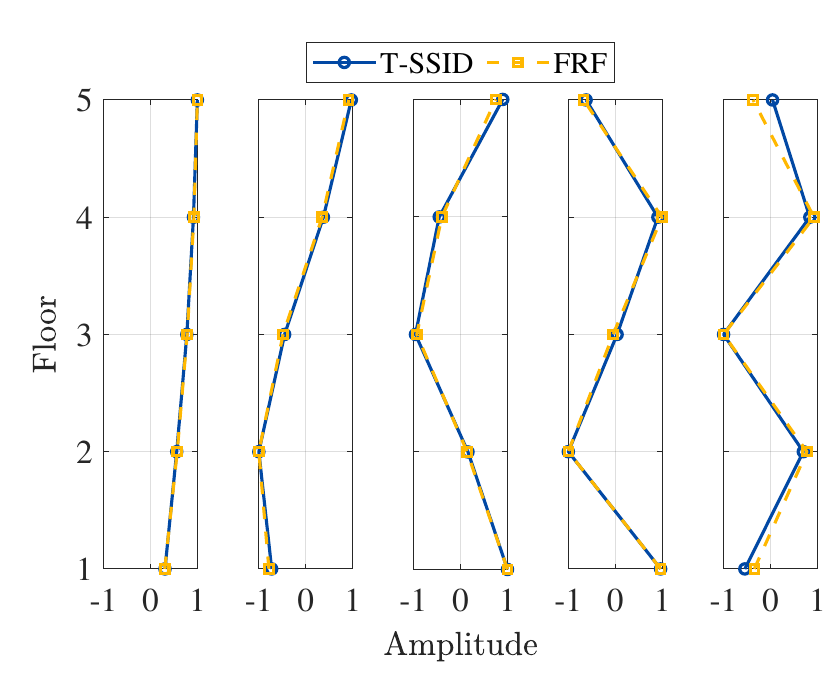}
    \caption{The identified mode shapes of the test structure.}
    \label{fig:mode_shapes}
\end{figure}
 
\section{Experimental implementation and results}
\label{section4}

\noindent In this section, experimental implementation and validation of the UF and US are presented. The objective is to assess their accuracy, robustness, and adaptability under various conditions through shake table and multi-impact tests introduced in Section~\ref{section3}. The performance of the proposed methods is evaluated against several state-of-the-art estimators, i.e., the AKF~\cite{LourensReyndersEATAL2012} and  GDFs~\cite{Gillijns2007a, Gillijns2007b}. In addition, the implementation details of the proposed methods, including initialisation strategies, adaptive tuning configurations, sensor placements for partial measurements, and the experimental arrangements used to ensure fair comparisons, are also described in this section. 

\subsection{Estimation setup and comparison framework}
\noindent To ensure consistent evaluation and fair comparison of the UF, US, and other benchmark estimators, the following initialisation and adaptive tuning setups were adopted across all experimental configurations.

For all estimation methods, the initial state vector $\hat{\mathbf{x}}_{0|0}$ is set to zero, and the associated error covariance $\mathbf{P}_0^{\mathbf{x}}$ is set to $1\times10^{-9}$, a small non-zero value to prevent numerical issues. The measurement noise covariance matrix $\mathbf{R}_k$ is defined based on the signal-to-noise ratio characteristics of the measurement systems, resulting in a constant matrix $\mathbf{R}$ used throughout the estimation process. The process error covariance matrix $\mathbf{Q}_k$ is assumed to be diagonal, expressed as $\mathbf{Q}_k=Q_k\mathbf{I}$, where the scalar $Q_k$ is tuned using the adaptive tuning scheme introduced in Section~\ref{section2_4}. 

In the adaptive tuning scheme, an array of candidate filters is deployed in parallel, each configured with a distinct value of $Q_k$. The values of $Q_k$ are sampled logarithmically from $10^{-24}$ to $10^{3}$, generating a total of $2,700$ candidate filters with a uniform spacing of $10^{0.01}$. At each timestep, the candidate filter that yielded the lowest accumulated innovation error $\mathbf{E}_k$ over a fixed-length evaluation window $W=10$ is selected. The corresponding state, input, and covariance estimates of the selected filter were used to reinitialise the filter array for the next cycle.

It should be noted that some specific settings are applied to the US and AKF. For the US, the smoothing time window is set to $N=20$ timesteps, and the initial cross-covariance matrices $\mathbf{P}_0^{\mathbf{xw}}$ and $\mathbf{P}_0^{\mathbf{xv}}$ are set to zero. For the AKF, which uses an augmented state vector, the corresponding process error covariance matrix is assumed as $\mathbf{Q}_k = \begin{bmatrix}Q_{k}^{x}\mathbf{I} & \mathbf{0} \\\mathbf{0} & Q_{k}^{p}\mathbf{I} \end{bmatrix}$, in which $Q_{k}^{x}$ represents the process error of the state and is searched from $10^{-24}$ to $10^{3}$, whereas $Q_{k}^{p}$ represents the process error of the fictitious evaluation model for the input and is searched from $10^{-18}$ to $10^{9}$. Since the AKF requires tuning of two independent parameters, the computational cost of applying the adaptive scheme grows substantially, making real-time or near-real-time operation impractical. Nonetheless, this is not a limitation for MVU-based methods, including UF, US and GDFs. To balance computational cost and estimation accuracy, the AKF in the array is set to $8,100$ candidate filters, resulting in a coarser sampling interval of $10^{0.3}$.

Finally, the estimation accuracy is evaluated using the Normalised Root Mean Square Error (NRMSE), defined as:
\begin{equation}
    \mathrm{NRMSE} = \sum_{i}^{t_{end}}\left[ \frac{\Delta \alpha_{rms,i}}{(\alpha_{max,i}-\alpha_{min,i})}\right],
\end{equation}

\noindent in which $\alpha$ represents the quantity being estimated (e.g., displacement, velocity, acceleration, or input); $\alpha_{rms,i}$ is the root-mean-square error at each timestep; $\alpha_{max.i}$ and $\alpha_{min,i}$ are the absolute maximum and minimum true values of $\alpha$, respectively. Detailed comparisons of the proposed methods with the benchmark estimators under different excitation scenarios are presented next.

\subsection{Ground motion results}

\noindent The performance of the proposed UF and US methods is evaluated under shake table excitation using three different sensor configurations, as presented in Table~\ref{tab:config_shake_table}. The goal is to assess the robustness and adaptability of the estimators across scenarios with and without direct feedthrough, as well as varying measurement types. Each configuration presents distinct challenges, reflecting practical sensor limitations encountered in structural monitoring applications. Configuration 1 features a typical setup with both displacement and acceleration measurements, forming a system with direct feedthrough. This configuration is used to establish a baseline performance for all estimators under favourable observability conditions. Configuration 2 represents a displacement-only system without direct feedthrough. This setup is motivated by computer vision-based monitoring systems, where typically only displacement data is available. Such configurations are known to amplify the effect of noise in the input estimation due to the absence of velocity and acceleration data~\cite{Mohsen2022, Liu2025}. The proposed US is expected to effectively mitigate this noise amplification. Configuration 3 comprises acceleration-only measurements. MVU-based estimators are known to suffer from a so-called drift effect under this setup, while the AKF tends to experience reduced accuracy. This configuration is particularly valuable for highlighting the advantages of the US.
\begin{table}[hb]
    \centering
    \small
    \caption{Measurement configurations for the ground motion scenario.}
    \label{tab:config_shake_table}
    \begin{tabular}{c c c}
    \toprule
     & \multicolumn{2}{c}{Location of sensors} \\
    \cmidrule{2-3}
    Configurations & Displacement & Acceleration \\
    \midrule
    1 & F1 & F2, F4 \\
    2 & F1, F4 & - \\
    3 & - & F3, F5 \\
    \bottomrule
    \end{tabular}
\end{table}

\begin{figure}[ht]
    \centering
    \begin{subfigure}{0.85\textwidth}
        \caption{} \label{fig:config1_input_time}
        \includegraphics[trim={2cm 7.1cm 2cm 7cm}, clip, width=\textwidth]{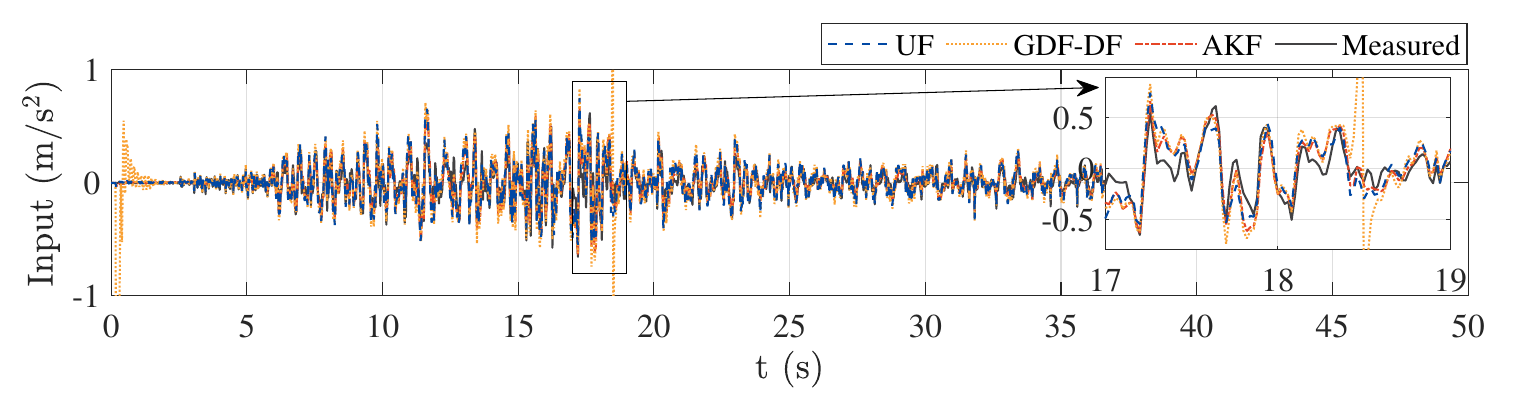}
    \end{subfigure}
    \begin{subfigure}{0.85\textwidth}
        \caption{} \label{fig:config1_input_fft}
        \includegraphics[trim={2cm 7.1cm 2cm 7cm}, clip, width=\textwidth]{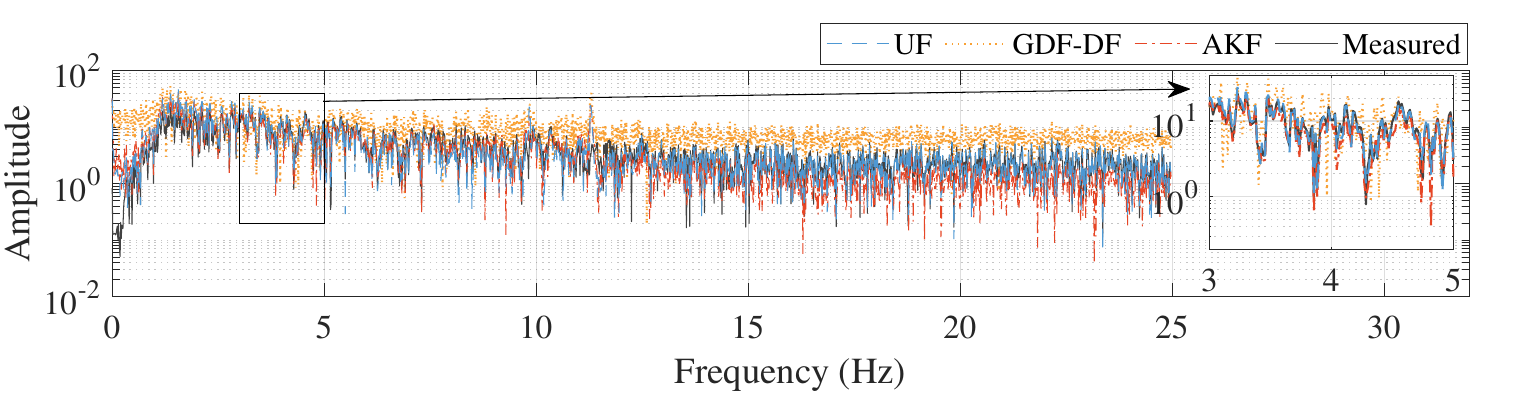}
    \end{subfigure}
    \caption{Input estimation of configuration 1: (a) time history, and (b) frequency spectrum.}
    \label{fig:config1_input}
\end{figure}

\begin{figure}[ht]
    \centering
    \begin{subfigure}{0.48\textwidth}
        \caption{} \label{fig:config1_disp}
        \includegraphics[trim={3cm 11.5cm 4cm 12cm}, clip, width=\textwidth]{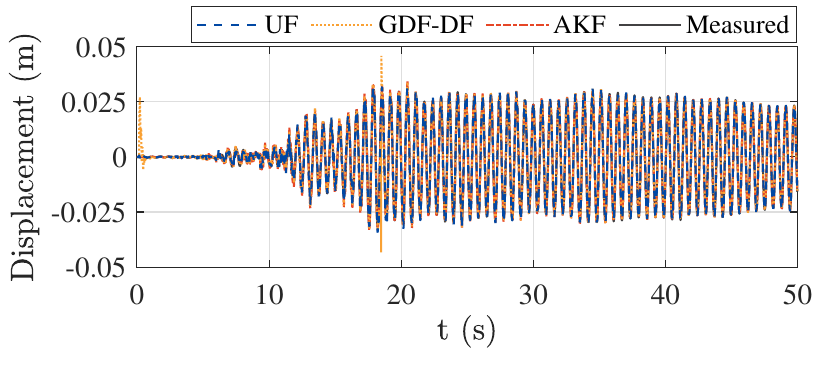}
    \end{subfigure}
    \begin{subfigure}{0.48\textwidth}
        \caption{} \label{fig:config1_accel}
        \includegraphics[trim={3cm 11.5cm 4cm 12cm}, clip, width=\textwidth]{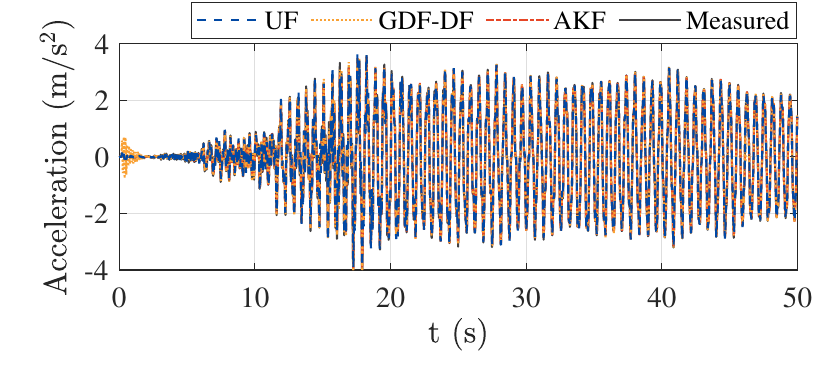}
    \end{subfigure}
    \caption{State estimation of configuration 1: (a) displacement and (b) acceleration.}
    \label{fig:config1_state}
\end{figure}

\clearpage
Under Configuration 1, the system has full-rank direct feedthrough, with displacement measured at Floor 1 and acceleration at floors 2 and 4. This sensor configuration provides strong observability for all estimators. Fig.~\ref{fig:config1_input} compares the input estimation performance of the UF against GDF-DF and AKF in both the time and frequency domains. The accuracy improvement introduced by smoothing is not significant, and therefore, the US results are not represented. As noticed in the time history, the GDF-DF exhibits strong fluctuations when the ground motion intensifies between $12$ and $20$~s, whereas the UF and AKF produce noticeably smoother estimates. This sensitivity to noise is more clearly illustrated in the frequency domain shown in Fig.~\ref{fig:config1_input_fft}, where the GDF-DF suffers from broadband noise amplification, and the AKF underestimates the input at higher frequencies. In contrast, the UF maintains a stable input estimation across the full spectrum. As a result, the NRMSE for input estimation is $0.078$ for the UF, compared to $0.086$ and $0.228$ of the AKF and GDF-DF, respectively. The state estimation results are shown in Fig.~\ref{fig:config1_state}. It should be noted that based on the state-space model defined by Eqs.~\eqref{eq:process} and ~\eqref{eq:measurement}, the acceleration estimation is not obtained directly. Instead, it is calculated by substituting the estimated state and input into the equation of motion: $\ddot{\mathbf{u}}_{k}=\mathbf{M}_{s}^{-1}\left( \mathbf{S}\mathbf{p}_k-\mathbf{C}_s\dot{\mathbf{u}}_k-\mathbf{K}_s\mathbf{u}_k \right)$. Nevertheless, all filters provide satisfactory state estimation, though there is a fluctuation in the displacement around $18$~s with the GDF-DF, which corresponds to the previously observed input error. 
\clearpage

\begin{figure}[ht]
    \centering
    \begin{subfigure}{0.85\textwidth}
        \caption{} \label{fig:config2_input_time}
        \includegraphics[trim={2cm 7.4cm 2cm 7.1cm}, clip, width=\textwidth]{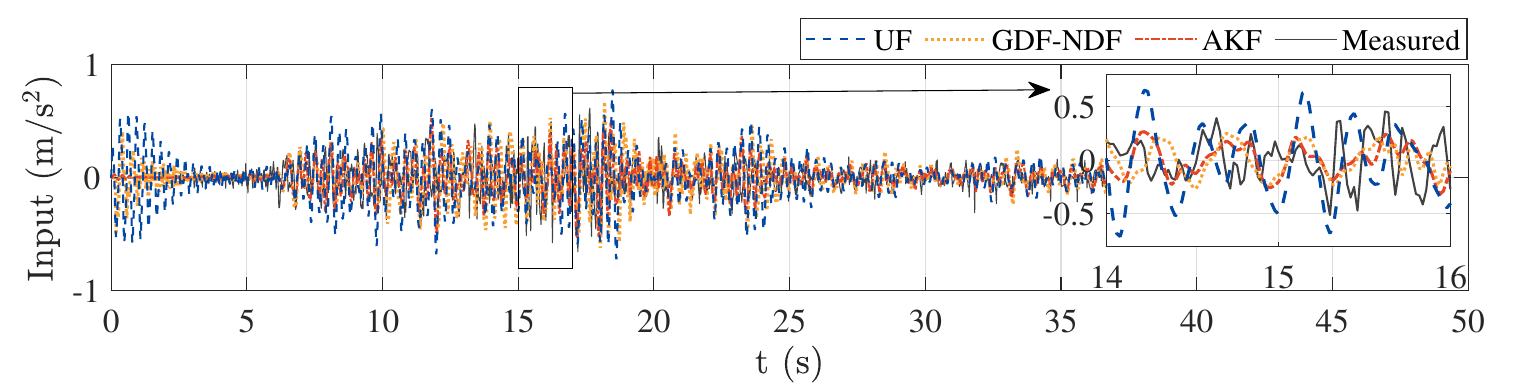}
    \end{subfigure}
    \begin{subfigure}{0.85\textwidth}
        \caption{} \label{fig:config2_input_sm}
        \includegraphics[trim={2cm 7.4cm 2cm 7.1cm}, clip, width=\textwidth]{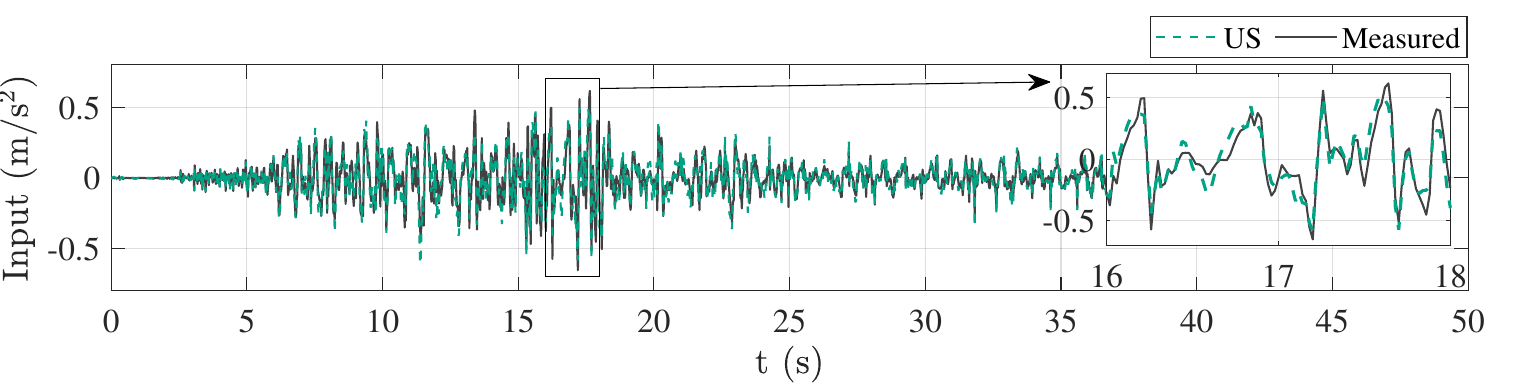}
    \end{subfigure}
    \begin{subfigure}{0.85\textwidth}
        \caption{} \label{fig:config2_input_fft}
        \includegraphics[trim={2cm 7.4cm 2cm 7.1cm}, clip, width=\textwidth]{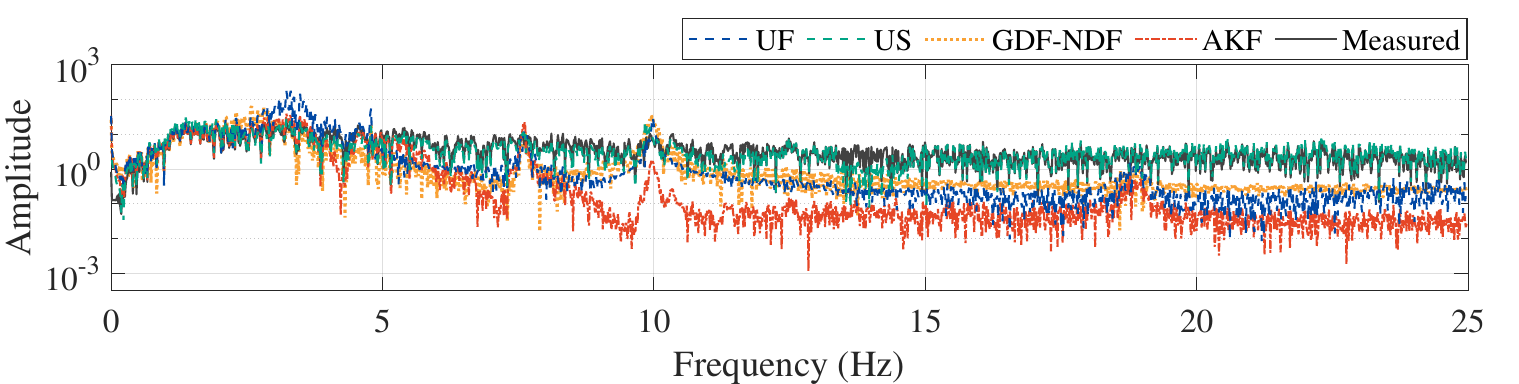}
    \end{subfigure}
    \caption{Input estimation of configuration 2: (a) time history by the UF, GDF-NDF and AKF, (b) time history by the US, and (c) frequency spectrum.}
    \label{fig:config2_input}
\end{figure}

\begin{figure}[ht]
    \centering
    \begin{subfigure}{0.9\textwidth}
        \caption{} \label{fig:config2_disp}
        \includegraphics[trim={2cm 7.9cm 2cm 7.5cm}, clip, width=\textwidth]{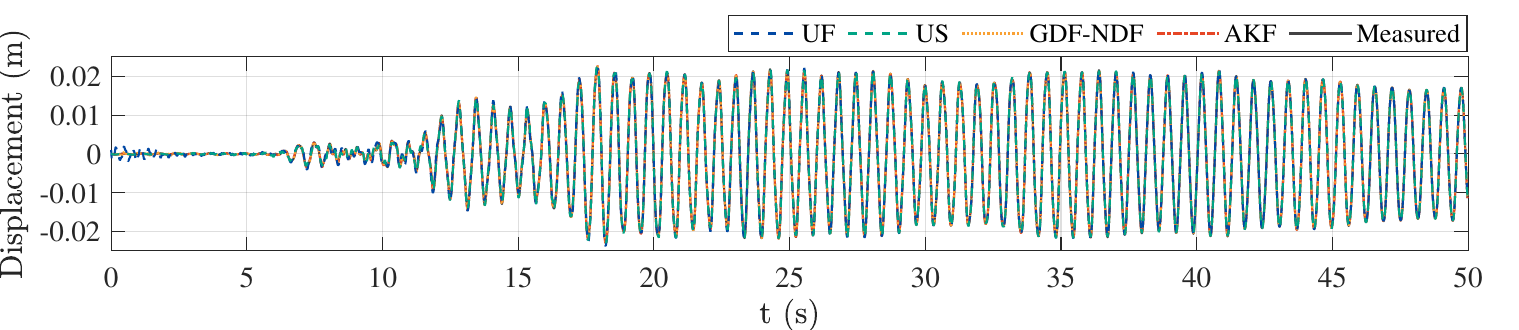}
    \end{subfigure}
    \begin{subfigure}{0.9\textwidth}
        \caption{} \label{fig:config2_accel}
        \includegraphics[trim={2cm 7.9cm 2cm 7.5cm}, clip, width=\textwidth]{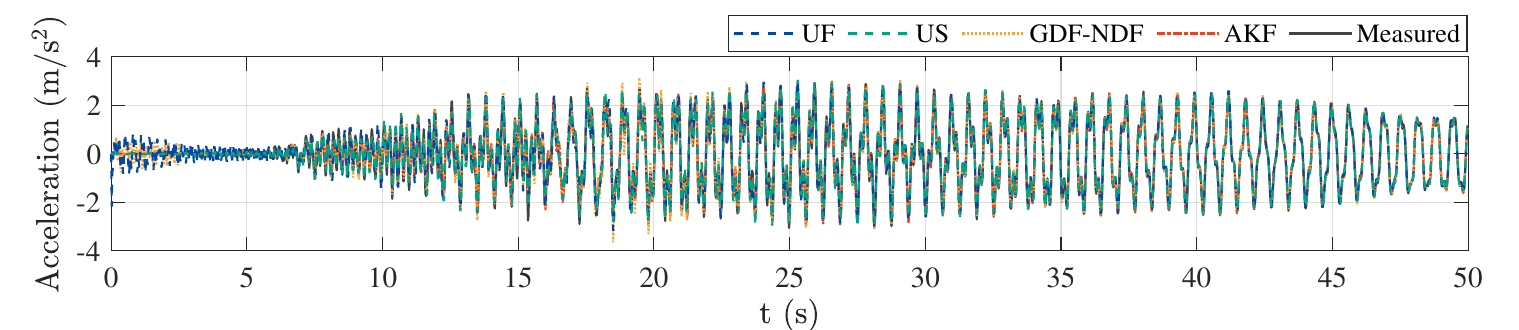}
    \end{subfigure}
    \caption{State estimation of configuration 2: (a) displacement, and (b) acceleration.}
    \label{fig:config2_state}
\end{figure}

Configuration 2 represents a displacement-only system, representing sensor layouts common in computer vision-based monitoring. In this setup, displacement is measured at floors 1, 4, and 5, and no acceleration or velocity sensors are available. As found in Fig.~\ref{fig:config2_input}, the absence of velocity and acceleration data significantly challenges the estimators. The input time histories reveal that both the AKF and GDF-NDF are heavily affected, with noisy oscillations dominating the estimates. Since the initial conditions are not specifically tuned, all filtering approaches take approximately 4 seconds to converge to stable estimates. The UF captures part of the vibration profile in the low-frequency range but fails to capture the mid- and high-frequency components, resulting in overestimation in the time domain. In contrast, the US -- which incorporates a 20-timestep observation window -- substantially improves the input estimation, visible in both the time domain and frequency spectrum. The US achieves a balance between noise suppression and dynamic fidelity, clearly having the best performance. State estimation results are presented in Fig.~\ref{fig:config2_state}. From Fig.~\ref{fig:config2_disp},  all methods achieve good matches in displacements; however, estimation errors of input and state accumulate in the filtering approaches, leading to fluctuating acceleration estimation shown in Fig.~\ref{fig:config2_accel}. The overall NRMSE is $0.438$ for the US and $0.953$ for the UF, compared to $0.605$ for the AKF and $0.562$ for the GDF-NDF. These results highlight the substantial benefit of the proposed smoothing method in displacement-only configurations, not only by improving observability under such sensor constraints but also by enabling practical integration with vision-based structural health monitoring and digital twin systems.

\begin{figure}[ht]
    \centering
    \begin{subfigure}{0.85\textwidth}
        \caption{} \label{fig:config3_input_time}
        \includegraphics[trim={2cm 7.4cm 2cm 7.1cm}, clip, width=\textwidth]{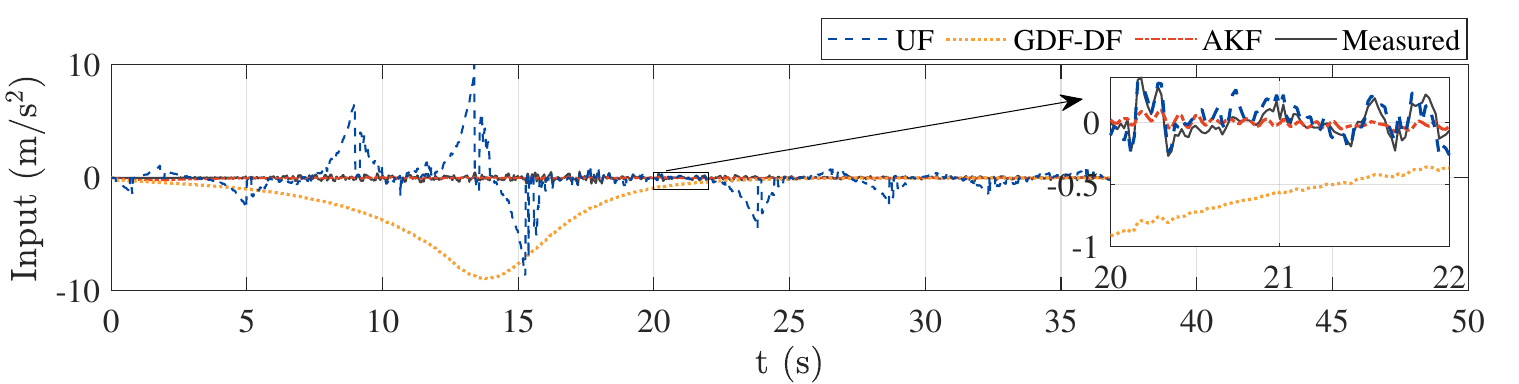}
    \end{subfigure}
    \begin{subfigure}{0.85\textwidth}
        \caption{} \label{fig:config3_input_sm}
        \includegraphics[trim={2cm 7.4cm 2cm 7.1cm}, clip, width=\textwidth]{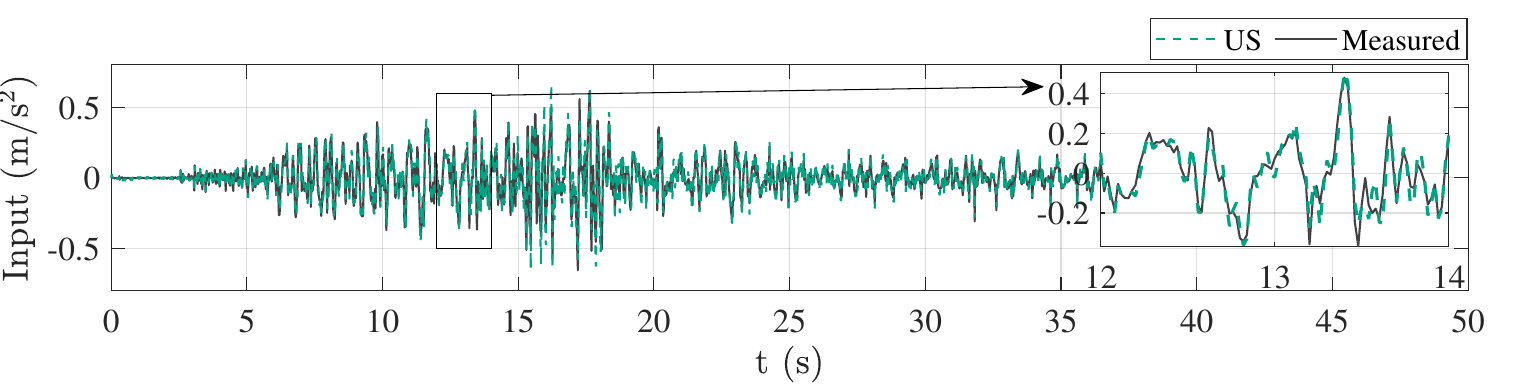}
    \end{subfigure}
    \begin{subfigure}{0.85\textwidth}
        \caption{} \label{fig:config3_input_fft}
        \includegraphics[trim={2cm 7.4cm 2cm 7.1cm}, clip, width=\textwidth]{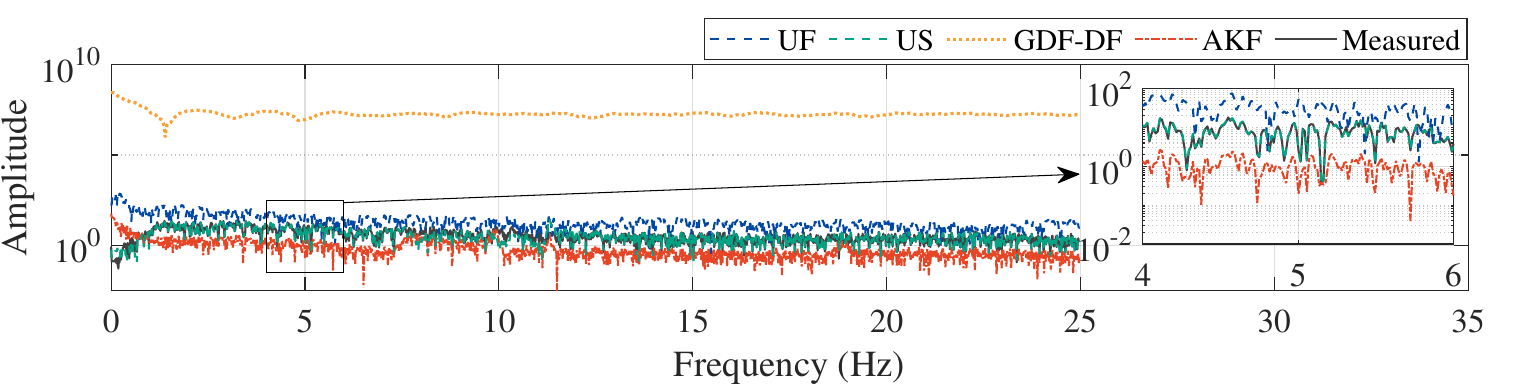}
    \end{subfigure}
    \caption{Input estimation of configuration 3: (a) time history by the UF, GDF-DF and AKF, (b) time history by the US, and (c) frequency spectrum.}
    \label{fig:config3_input}
\end{figure}

\begin{figure}[ht]
    \centering
    \begin{subfigure}{0.85\textwidth}
        \caption{} \label{fig:config3_disp}
        \includegraphics[trim={2cm 7.4cm 2cm 7.1cm}, clip, width=\textwidth]{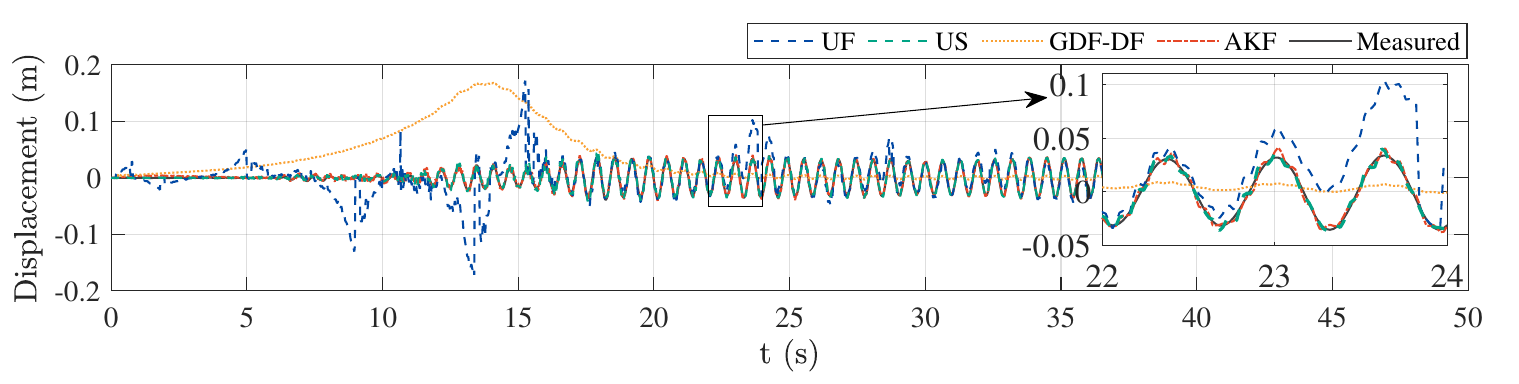}
    \end{subfigure}
    \begin{subfigure}{0.85\textwidth}
        \caption{} \label{fig:config3_accel}
        \includegraphics[trim={2cm 7.4cm 2cm 7.1cm}, clip, width=\textwidth]{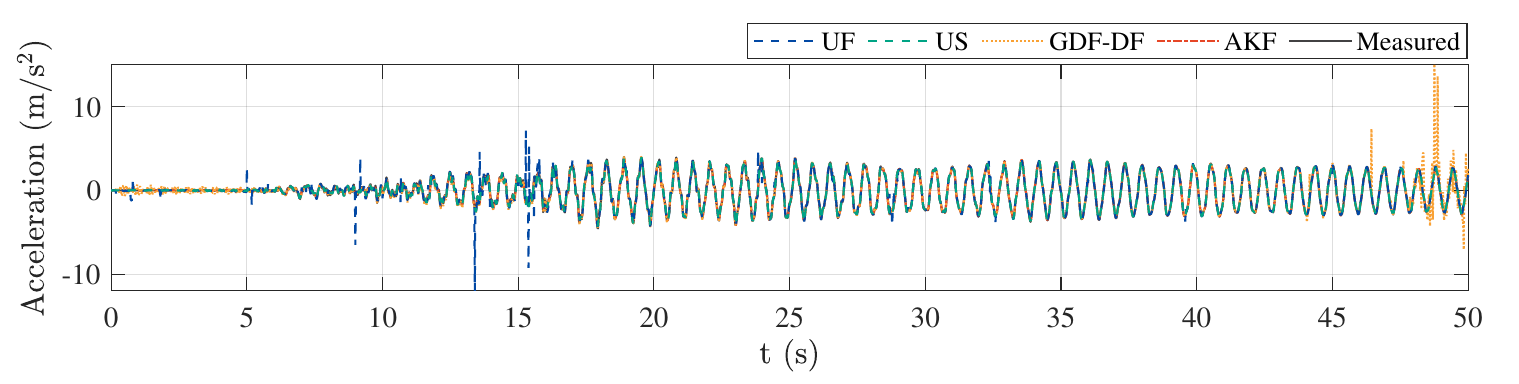}
    \end{subfigure}
    \caption{State estimation of configuration 3: (a) displacement, and (b) acceleration.}
    \label{fig:config3_state}
\end{figure}

Configuration 3 is an acceleration-only system, where acceleration is measured at floors~3 and 5, and no displacement or velocity data are available. This configuration poses significant challenges for input-state estimation, particularly for MVU-based methods. From Fig.~\ref{fig:config3_input}, the GDF-DF exhibits a clear drift in the input estimation, which accumulates over time and leads to a substantial deviation from the measured ground motion. This is a known limitation of MVU-based filters under acceleration-only setups, where the lack of displacement-level measurements allows low-frequency errors to propagate unbounded. The AKF, while more stable, consistently underestimates the input and fails to capture the full amplitude of excitation. The UF is able to capture some of the vibrational characteristics, particularly in the $16$-$22$~s; however, its input estimate is distorted by low-frequency fluctuations stemming from limited observability and the absence of displacement correction. In contrast, the US, seen in Fig.~\ref{fig:config3_input_sm}, shows improved performance by effectively suppressing the drift issue through the 20-timestep observation window. As illustrated in Fig.~\ref{fig:config3_state}, the AKF shows improvement in the state estimation, and both displacement and acceleration results are satisfactory. The UF shows a lighter form of drift in the displacement results, with most vibration profiles well captured and acceleration estimation remaining acceptable. In comparison, the US provides accurate and stable estimates that closely match the measured responses, suggesting an advantage in handling acceleration-only configurations. For reference, the overall NRMSE is $0.489$ for the US, compared to $0.524$ for the AKF, $5.048$ for the UF, and $30.131$ for the GDF-DF.

\subsection{Multi-impact results}

\noindent The multi-impact test results are presented herein to validate one of the key strengths of the proposed UF and US: their applicability to systems with a rank-deficient feedforward matrix. Such configurations are often encountered in practical scenarios involving limited or suboptimal sensor placement, which violate the full-rank conditions required by most existing MVU methods, such as~\cite{Gillijns2007b, Maes2018}. Unlike traditional estimators that are restricted to either full-rank or no-feedthrough systems, the UF and US can seamlessly handle rank-deficient conditions without requiring fictitious input models or statistical assumptions.

Two sensor configurations used for the multi-impact tests are summarised in Table~\ref{tab:config_impact}. Both configurations include only two acceleration sensors, while the number of unknown inputs (i.e., the hammer impacts) is three. As a result, the corresponding feedforward matrix $\mathbf{D}$ becomes rank-deficient, presenting a scenario where other MVU-based estimators are inapplicable. Configuration 5 features collocated sensors at the floors where the impacts are applied. This setup is used to evaluate whether the proposed estimators can resolve the inputs despite the lack of full-rank conditions. Configuration 6 introduces additional complexity by placing one of the sensors at a floor that is not directly excited. This non-collocated configuration challenges the estimators to detect and correctly attribute inputs that are not directly sensed. As the rank-deficient nature of the feedforward matrix falls outside the applicability of most conventional MVU-based methods, only the results obtained from the UF and US are presented in this section for comparison.
\begin{table}[ht!]
    \centering
    \small
    \caption{Measurement configurations for the multi-impact scenario.}
    \label{tab:config_impact}
    \begin{tabular}{c c c}
    \toprule
     & \multicolumn{2}{c}{Location of sensors} \\
    \cmidrule{2-3}
    Configurations & Displacement & Acceleration \\
    \midrule
    5 & F2  & F4, F5 \\
    6 & F1 & F4, F5 \\
    \bottomrule
    \end{tabular}
\end{table}

\begin{figure}[ht]
    \centering
    \begin{subfigure}{0.49\textwidth}
        \caption{} \label{fig:Figures/Impacts/impact_config1_input1.pdf}
        \includegraphics[trim={3cm 10.75cm 4cm 11cm}, clip, width=\textwidth]{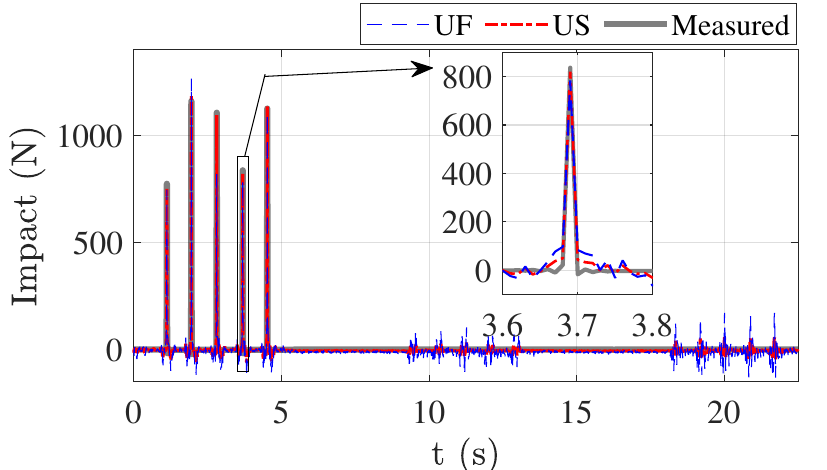}
    \end{subfigure}
    \begin{subfigure}{0.49\textwidth}
        \caption{} \label{fig:impact_config1_input2}
        \includegraphics[trim={3cm 10.75cm 4cm 11cm}, clip, width=\textwidth]{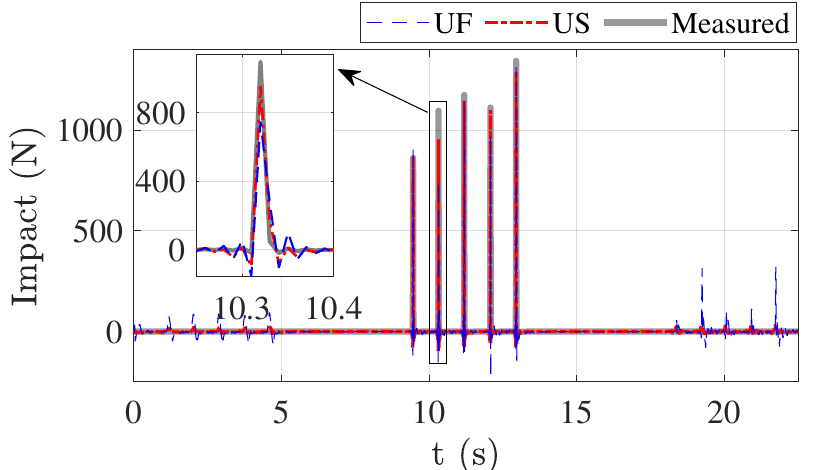}
    \end{subfigure}
    \begin{subfigure}{0.49\textwidth}
        \caption{} \label{fig:impact_config1_input3}
        \includegraphics[trim={3cm 10.75cm 4cm 11cm}, clip, width=\textwidth]{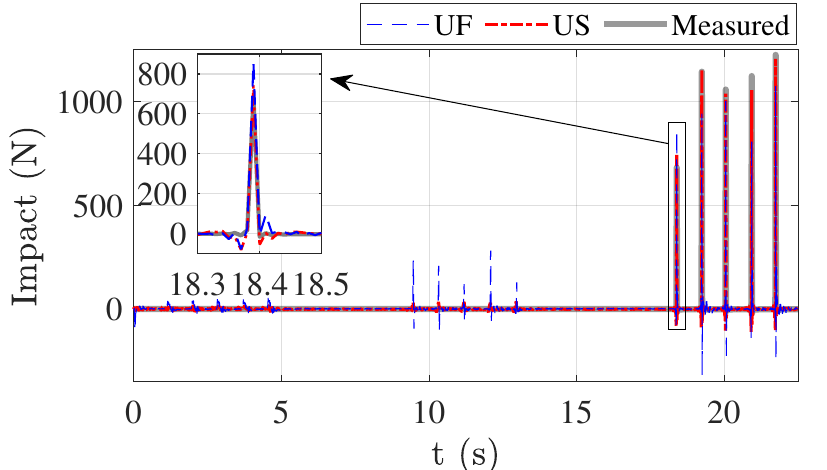}
    \end{subfigure}
    \caption{Input estimation results of configuration 5: (a) the first impact on floor 1, (b) the second impact on floor 5, and (c) the third impact on floor 4.}
    \label{fig:impact_config1_input}
\end{figure}

\begin{figure}[ht]
    \centering
    \begin{subfigure}{0.85\textwidth}
        \caption{} \label{fig:impact_config1_disp}
        \includegraphics[trim={2cm 7.4cm 2cm 7.1cm}, clip, width=\textwidth]{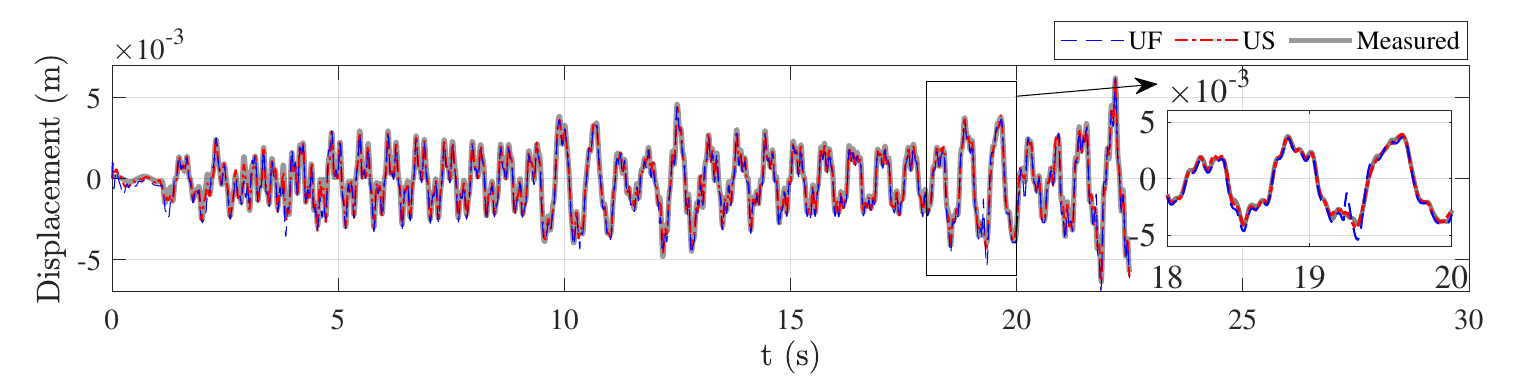}
    \end{subfigure}
    \begin{subfigure}{0.85\textwidth}
        \caption{} \label{fig:impact_config1_accel}
        \includegraphics[trim={2cm 7.4cm 2cm 7.1cm}, clip, width=\textwidth]{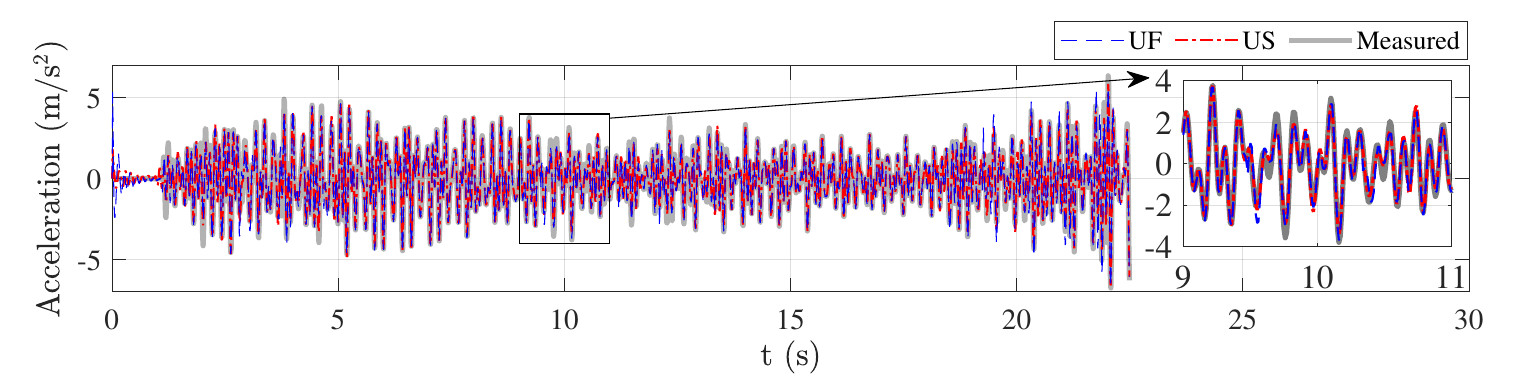}
    \end{subfigure}
    \caption{State estimation results of configuration 5: (a) displacement and (b) acceleration. }
    \label{fig:impact_config1_state}
\end{figure}

Configuration 5 consists of two acceleration and one displacement measurements, and all are collocated with the applied hammer impacts at floors 2, 4 and 5. Since there are three unknown inputs but only two acceleration measurements, the resulting feedforward matrix is rank-deficient. However, the system inversion requirement by the proposed universal methods is still satisfied, i.e., the total number of measurements of any type is greater or equal to the number of unknown inputs. As represented in Fig.~\ref{fig:impact_config1_input}, the UF captures the dominant peaks in the inputs corresponding to the floor impacts but underestimates the relative magnitudes of some events, such as the third tap on floor 2 and the second tap on floor 5. Residual noise is also observed during intervals when other impacts occur. In contrast, the US achieves a clearer and more accurate reconstruction of the three input events, with clearly reduced high-frequency fluctuations. Fig.\ref{fig:impact_config1_state} illustrates the corresponding state estimations. Both methods provide satisfactory performance in estimating floor displacements and accelerations. However, the UF tends to show slight fluctuation following each impact due to the sudden change in the input but recovers quickly. The US maintains consistency across the entire time history, confirming enhanced performance due to the extended observation window. The overall NRMSE is $0.655$ of the UF and $0.604$ of the US.

\begin{figure}[ht]
    \centering
    \begin{subfigure}{0.49\textwidth}
        \caption{} \label{fig:Figures/Impacts/impact_config2_input1.pdf}
        \includegraphics[trim={3cm 10.75cm 4cm 11cm}, clip, width=\textwidth]{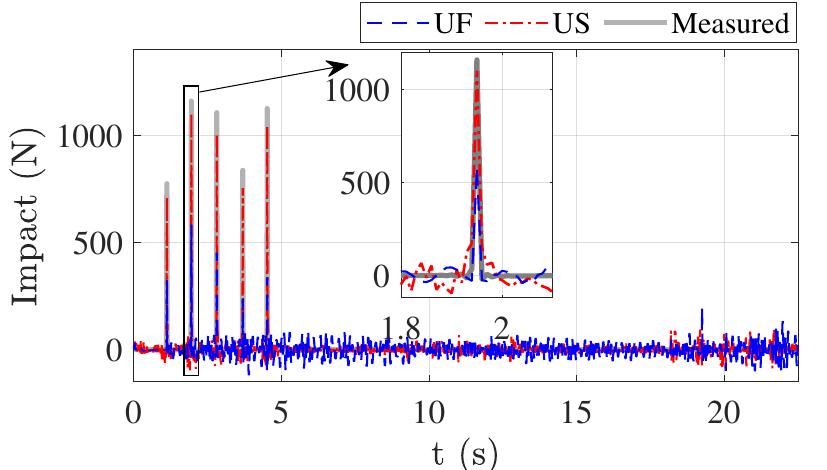}
    \end{subfigure}
    \begin{subfigure}{0.49\textwidth}
        \caption{} \label{fig:impact_config2_input2}
        \includegraphics[trim={3cm 10.75cm 4cm 11cm}, clip, width=\textwidth]{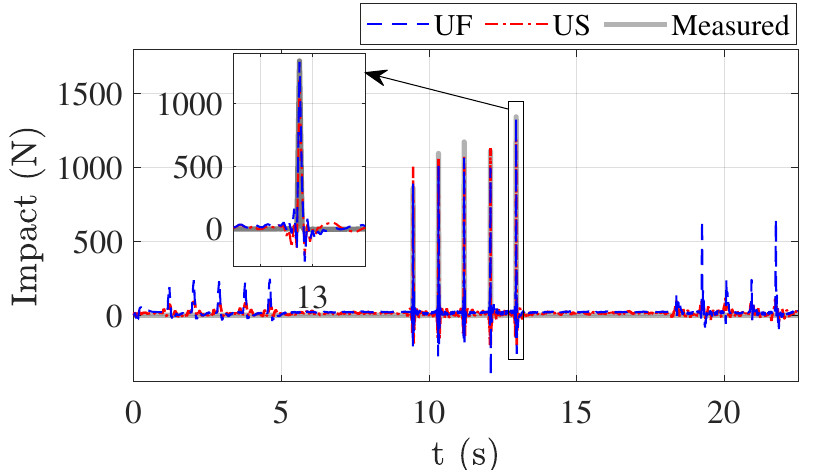}
    \end{subfigure}
    \begin{subfigure}{0.49\textwidth}
        \caption{} \label{fig:impact_config2_input3}
        \includegraphics[trim={3cm 10.75cm 4cm 11cm}, clip, width=\textwidth]{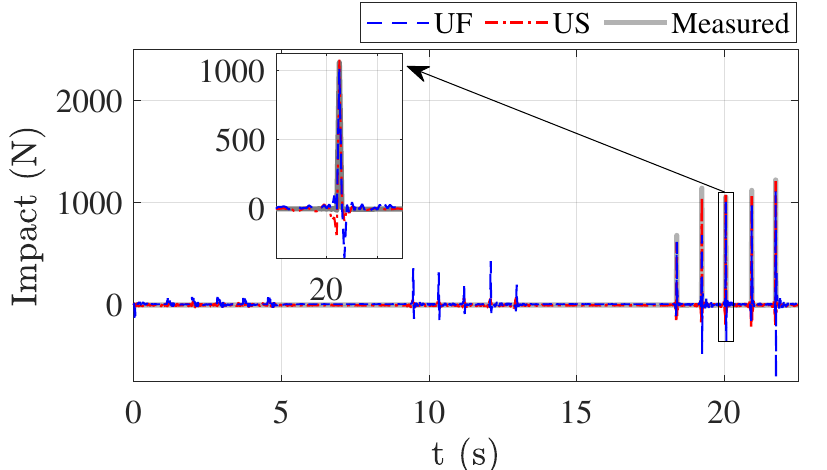}
    \end{subfigure}
    \caption{Input estimation results of configuration 6: (a) the first impact on floor 1, (b) the second impact on floor 5, and (c) the third impact on floor 4.}
    \label{fig:impact_config2_input}
\end{figure}

\begin{figure}[ht]
    \centering
    \begin{subfigure}{0.85\textwidth}
        \caption{} \label{fig:impact_config2_disp}
        \includegraphics[trim={2cm 7.4cm 2cm 7.1cm}, clip, width=\textwidth]{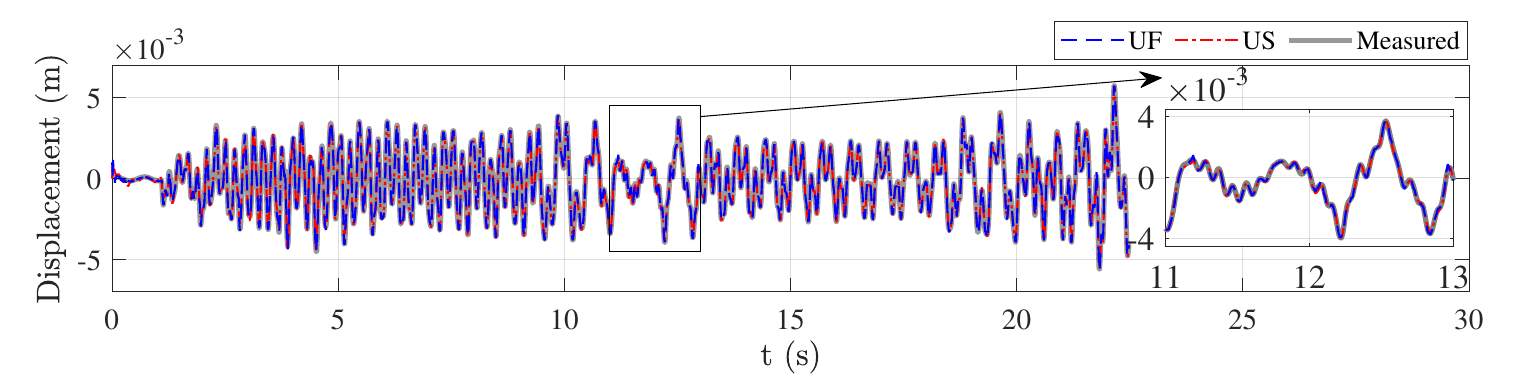}
    \end{subfigure}
    \begin{subfigure}{0.85\textwidth}
        \caption{} \label{fig:impact_config2_accel}
        \includegraphics[trim={2cm 7.4cm 2cm 7.1cm}, clip, width=\textwidth]{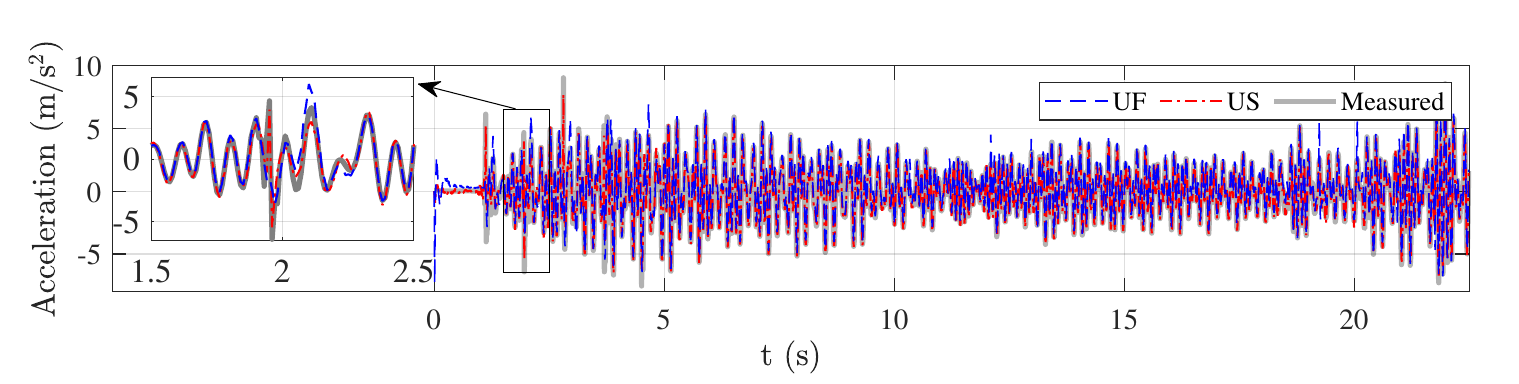}
    \end{subfigure}
    \caption{State estimation results of configuration 6: (a) displacement and (b) acceleration.}
    \label{fig:impact_config2_state}
\end{figure}

Configuration 6 presents a more challenging setup, where the single displacement measurement is not collocated with the applied hammer impacts on floor 2 but instead placed on the unexcited floor 1. Non-collocated measurements are known to pose difficulties for filtering-based estimators. As expected, the UF exhibits degraded input estimation accuracy for the impact applied to floor 2, as shown in Fig.\ref{fig:impact_config2_input}. The magnitudes of the impacts are underestimated, and the residual noise is intensified. It should be denoted that the noise displays a pattern that resembles acceleration responses. In contrast, the US continues to provide a robust performance, adequately estimating the timing and relative amplitudes of the impacts, with smoother and more stable input reconstruction. As a result, the NRMSE of the input estimation is $0.126$ for the UF and $0.093$ for the US. The state estimation results of the unmeasured floor 2 are presented in Fig.~\ref{fig:impact_config2_state}. Both methods provide reasonable displacement estimates; however, due to the compromised input accuracy of the UF, the resulting acceleration response lacks precision around the impact events. The US maintains consistent and reliable performance in both displacement and acceleration estimates, with minimal indication of drift or noise amplification. These results confirm the advantage of the smoothing approach in handling non-collocated systems where instantaneous filters begin to degrade. 

\clearpage
\subsection{Adaptive tuning performance and convergence}
\noindent The performance and convergence behaviour of the adaptive tuning scheme introduced in Section~\ref{section2_4} are examined. The proposed filter array approach enables the online selection of process noise covariance values without relying on offline tuning, offering practical advantages for real-time structural health monitoring. Here, the evolution of the selected values over time is analysed for representative test cases. In addition, the convergence of the a posteriori error covariance matrix is investigated to assess the numerical stability and consistency of the proposed methods. The results suggest that the adaptive tuning scheme successfully tracks changing input conditions while maintaining stable estimation error bounds across both ground motion and multi-impact scenarios.

\begin{figure}[ht]
    \centering
    \begin{subfigure}{0.49\textwidth}
        \caption{} \label{fig:eq_q_history}
        \includegraphics[trim={7.5cm 6cm 8.5cm 7cm}, clip, width=\textwidth]{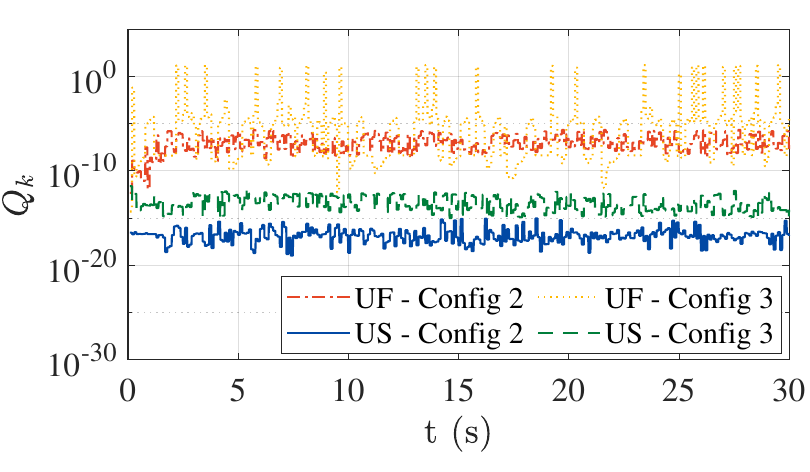}
    \end{subfigure}
    \begin{subfigure}{0.49\textwidth}
        \caption{} \label{fig:impact_q_history}
        \includegraphics[trim={7.5cm 6cm 8.5cm 7cm}, clip, width=\textwidth]{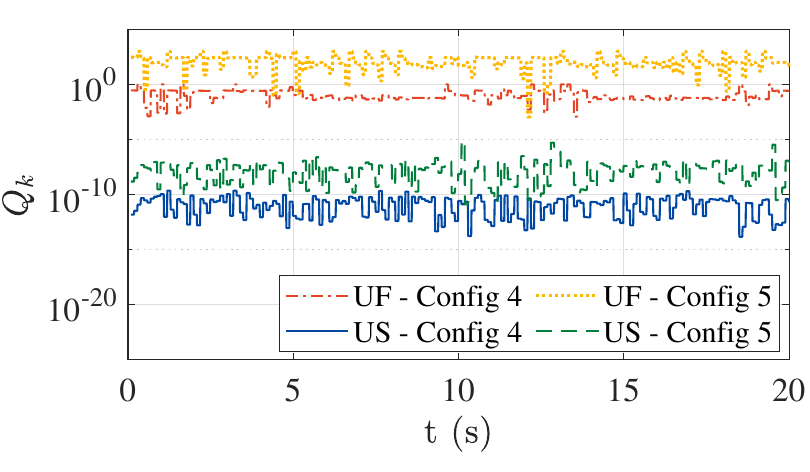}
    \end{subfigure}
    \caption{Time history of adaptive tuning: (a) the shake table test, and (b) the multi-impact test.}
    \label{fig:q_history}
\end{figure}

The evolution of the process noise covariance in the displacement-only and acceleration-only configurations (i.e., configurations 2 and 3) during the shake table tests is plotted in Fig.\ref{fig:eq_q_history}. For the displacement-only setup (configuration 2), both the UF and US adapt their process noise covariance within a stable range, although the UF generally selects higher values than the US. However, under the acceleration-only setup (configuration 3), the process noise covariance for the UF varies significantly, suggesting that candidates in the filter array are more strongly affected by the drift effect and corresponding large innovations. In contrast, the US maintains consistent and appropriate $Q_k$ selections across time, demonstrating improved stability. Fig.~\ref{fig:impact_q_history} shows the evolution of $Q_k$ during the multi-impact tests. The tuning mechanism successfully tracks varying dynamic conditions by adjusting $Q_k$ selections in response to different loading events. This behaviour confirms that the proposed adaptive scheme provides real-time flexibility and robustness without requiring manual intervention or prior assumptions about the noise environment.

\begin{figure}[ht]
    \centering
    \begin{subfigure}{0.45\textwidth}
        \caption{} \label{fig:eq_cov}
        \includegraphics[trim={3cm 10.5cm 4.2cm 11.5cm}, clip, width=\textwidth]{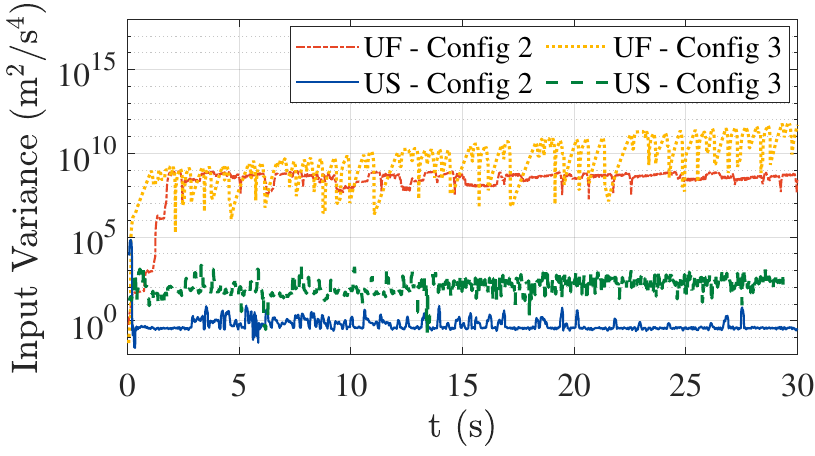}
    \end{subfigure}\\
    \begin{subfigure}{0.32\textwidth}
        \caption{} \label{fig:inpmact_cov_1}
        \includegraphics[trim={3cm 9.4cm 4.5cm 10cm}, clip, width=\textwidth]{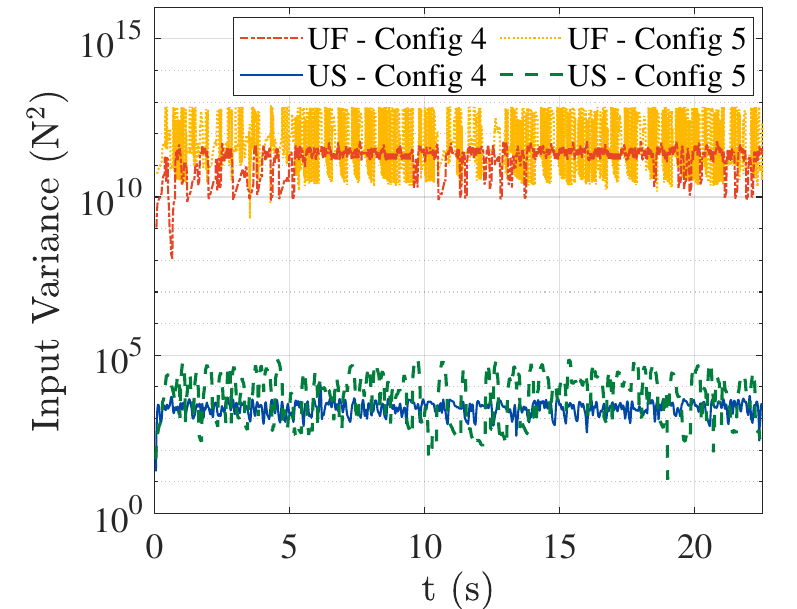}
    \end{subfigure}
    \begin{subfigure}{0.32\textwidth}
        \caption{} \label{fig:inpmact_cov_2}
        \includegraphics[trim={3cm 9.4cm 4.5cm 10cm}, clip, width=\textwidth]{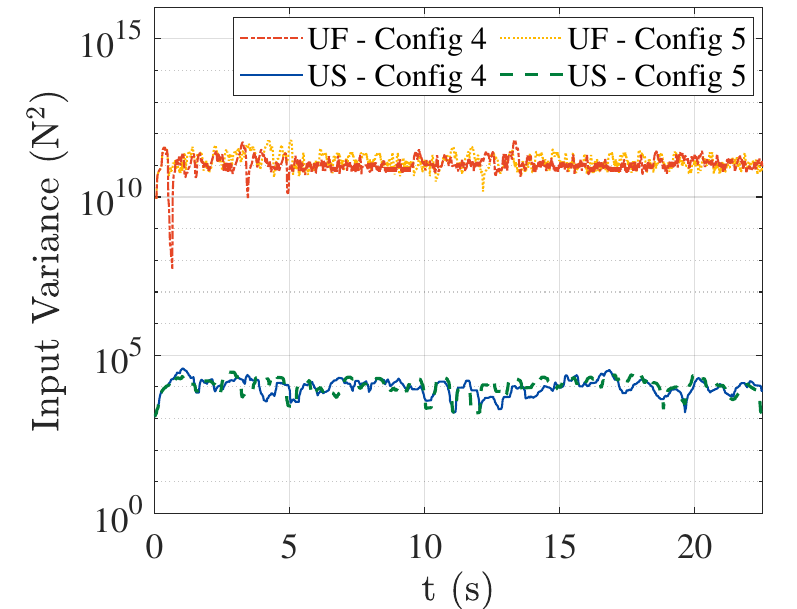}
    \end{subfigure}
    \begin{subfigure}{0.32\textwidth}
        \caption{} \label{fig:inpmact_cov_3}
        \includegraphics[trim={3cm 9.4cm 4.5cm 10cm}, clip, width=\textwidth]{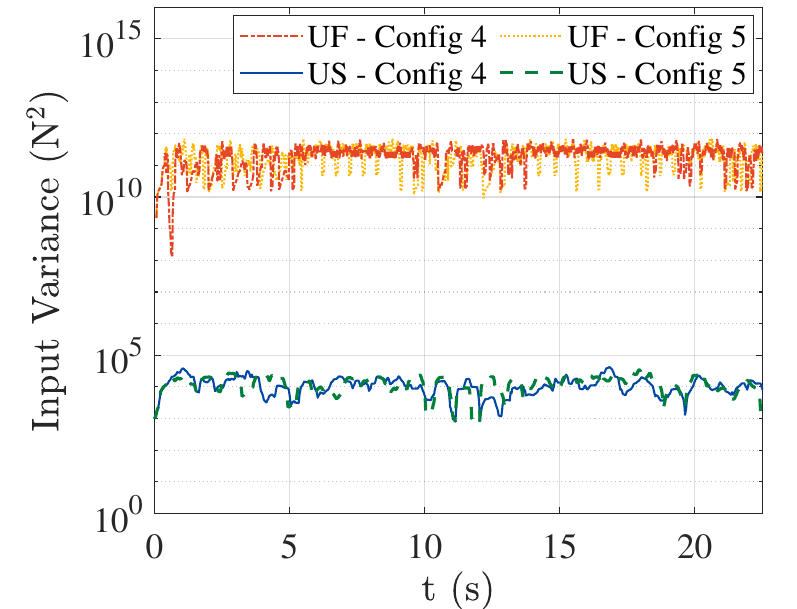}
    \end{subfigure}
    \caption{Error variance: (a) ground motion, (b) the impacts on the 2nd floor, (c) the impacts on the 5th floor, and (d) the impacts on the 4th floor. }
    \label{fig:cov_history}
\end{figure}

The evolution of the variances associated with the input estimation for the shake table test is presented in Fig.~\ref{fig:eq_cov}. It is evident that the error covariance of the UF under acceleration-only measurements (configuration 3) grows over time and fails to converge, ultimately leading to drift in the input estimation. In contrast, the US achieves convergence, although the variance remains higher than in the displacement-only configuration. It should be noted that fluctuations in the variance profiles are primarily caused by shifts between optimal candidates within the adaptive tuning array. The variances associated with each reconstructed impact force in the multi-impact test are plotted in Figs.\ref{fig:inpmact_cov_1} to\ref{fig:inpmact_cov_3}. Despite the rank-deficient feedforward matrix, both the UF and US operate in well-conditioned systems and show convergence without numerical issues. It should be noticed that in configuration 5, where no sensor is collocated with the impact force on the second floor, the corresponding variance exhibits increased noise, particularly in the UF. Nonetheless, convergence is still attained within a reasonable range comparable to collocated configurations (e.g., Configuration~4). Across all scenarios, the US maintains a lower and stable input error covariance. This reduction and stabilisation are direct outcomes of the smoothing mechanism, which takes advantage of the extra sensor information in the extended observation window to mitigate the effects of instantaneous measurement noise and modelling discrepancies. These findings confirm that, with a slight sacrifice of real-time capability (a 0.2~s delay in this study), the smoothing framework not only improves estimation accuracy but also substantially enhances estimation stability and robustness, particularly under challenging measurement conditions such as displacement-only and acceleration-only setups.

\section{Conclusions}
\label{section5}

\noindent This paper aimed to experimentally validate the Universal Filter and Universal Smoothing methods, originally developed for joint input and state estimations without restrictive sensor network requirements as typical of minimum-variance unbiased methods. In addition, an adaptive tuning scheme was developed to replace offline tuning procedures and enable real-time adaptability. A five-storey shear frame structure was tested under both ground motion and multi-impact excitations, incorporating a variety of sensor configurations, including scenarios with and without direct feedthrough as well as rank-deficient feedforward matrices. The key findings from the experimental validations are summarised below:

\begin{itemize}
\item In the ground motion tests, both the UF and US achieved higher estimation accuracy than existing methods, including the AKF and GDFs, across various sensor configurations, such as displacement-only and acceleration-only setups. The UF provided stable and accurate input-state estimation under typical measurement conditions, while the US further extended the robustness by utilising an extended observation window even under challenging scenarios with limited or noisy measurements.

\item In the multi-impact tests, which featured intentionally rank-deficient feedforward matrices, both the UF and US successfully achieved convergence without numerical issues, extending the applicability of the proposed framework to conditions where traditional MVU-based methods are typically inapplicable. The US, by incorporating an extended observation window, enhanced input reconstruction accuracy in the non-collocated measurement setups.

\item Regarding adaptive tuning performance and convergence, the proposed filter array scheme reliably tracked dynamic changes by adaptively selecting appropriate process noise covariances in real-time. The tuning mechanism successfully maintained estimation stability without manual intervention, while the UF and US consistently yielded robust and stable joint input-state estimation. Across all configurations, the US exhibited enhanced variance control through its smoothing strategy, contributing further to estimation reliability under dynamic conditions.
\end{itemize}

The novelty of this study lies in the first experimental validation of the Universal Filter and Universal Smoothing methods, with adaptive tuning achieved by filter arrays. The validated methods offer strong potential for real-world deployment in structural health monitoring, particularly in settings where sensor layouts are limited, unplanned, or evolve over time, such as infrastructures in hazardous environments, offshore platforms, or vision-based monitoring systems.

Future research directions could explore the extension of these methods to nonlinear structural systems and field deployment on full-scale civil structures subjected to operational or extreme loads.

\clearpage
\appendix
\section{Extended state-space matrices used in the smoothing method} 
\label{AppendixA}
\noindent The expression of extended vectors, $\mathbcal{y}_k$, $\mathbcal{p}_k$, $\mathbcal{w}_k$ and $\mathbcal{v}_k$ are defined below:
\begin{equation} \label{appendix_eq1}
    \mathbcal{y}_k\triangleq\begin{bmatrix}\mathbf{y}_k^T & \cdots & \mathbf{y}_{k+N}^T \\\end{bmatrix}^T,
\end{equation}
\vspace{-0.5cm}
\begin{equation} \label{appendix_eq2}
    \mathbcal{p}_k\triangleq\begin{bmatrix}\mathbf{p}_k^T & \cdots & \mathbf{p}_{k+N}^T \\\end{bmatrix}^T,
\end{equation}
\vspace{-0.5cm}
\begin{equation} \label{appendix_eq3}
    \mathbcal{w}_k\triangleq\begin{bmatrix}\mathbf{w}_{k-1}^T & \cdots & \mathbf{w}_{k+N-1}^T\\\end{bmatrix}^T,
\end{equation}
\vspace{-0.5cm}
\begin{equation} \label{appendix_eq4}
    \mathbcal{v}_k\triangleq\begin{bmatrix}\mathbf{v}_k^T & \cdots & \mathbf{v}_{k+N}^T\\\end{bmatrix}^T.
\end{equation}

In addition, the extended state-space matrices $\mathbcal{C}_k$, $\mathbcal{D}_k$, and $\mathbcal{H}_k$ in the extended observation equation, Eq.~\eqref{eq:extended_measurement}, are expressed below: 
\begin{equation} \label{appendix_eq5}
    \mathbcal{C}_k=
    \begin{bmatrix}\mathbf{C}_k\\
    \mathbf{C}_{k+1}\mathbf{A}_k\\
    \mathbf{C}_{k+2}\mathbf{A}_{k+1}\mathbf{A}_k\\
    \vdots\\
    \mathbf{C}_{k+N}\prod_{i=0}^{N-1}\mathbf{A}_{k+i}\\\end{bmatrix},
\end{equation}
\vspace{-11pt}
{\footnotesize
\begin{equation} \label{appendix_eq6}
    \mathbcal{D}_k=
    \begin{bmatrix}\mathbf{D}_k&\mathbf{0}&\mathbf{0}&\cdots&\mathbf{0}\\
    \mathbf{0}&\mathbf{C}_{k+1}\mathbf{B}_k+\mathbf{D}_{k+1}&\mathbf{0}&\cdots&\mathbf{0}\\
    \mathbf{0}&\mathbf{C}_{k+2}\mathbf{A}_{k+1}\mathbf{B}_k&\mathbf{C}_{k+2}\mathbf{B}_{k+1}+\mathbf{D}_{k+2}&\cdots&\mathbf{0}\\
    \mathbf{0}&\mathbf{C}_{k+3}\mathbf{A}_{k+2}\mathbf{A}_{k+1}\mathbf{B}_k&\mathbf{C}_{k+3}\mathbf{A}_{k+2}\mathbf{B}_{k+1}&\cdots&\mathbf{0}\\
    \vdots&\vdots&\vdots&\ddots&\vdots\\
    \mathbf{0}&\mathbf{C}_{k+N}\left(\prod_{i=1}^{N-1}\mathbf{A}_{k+i}\right)\mathbf{B}_k&\mathbf{C}_{k+N}\left(\prod_{i=2}^{N-1}\mathbf{A}_{k+i}\right)\mathbf{B}_{k+1}&\cdots&\mathbf{C}_{k+N}\mathbf{B}_{k+N-1}+\mathbf{D}_{k+N}\\\end{bmatrix},
\end{equation}
}
\vspace{-11pt}
\begin{equation} \label{appendix_eq7}
    \mathbcal{H}_k=
    \begin{bmatrix}\mathbf{0}&\mathbf{0}&\mathbf{0}&\cdots&\mathbf{0}\\
    \mathbf{0}&\mathbf{C}_{k+1}&\mathbf{0}&\cdots&\mathbf{0}\\    \mathbf{0}&\mathbf{C}_{k+2}\mathbf{A}_{k+1}&\mathbf{C}_{k+2}&\cdots&\mathbf{0}\\ \mathbf{0}&\mathbf{C}_{k+3}\mathbf{A}_{k+2}\mathbf{A}_{k+1}&\mathbf{C}_{k+3}\mathbf{A}_{k+2}&\cdots&\mathbf{0}\\
    \vdots&\vdots&\vdots&\ddots&\vdots\\
    \mathbf{0}&\mathbf{C}_{k+N}\prod_{i=1}^{N-1}\mathbf{A}_{k+i}&\mathbf{C}_{k+N}\prod_{i=2}^{N-1}\mathbf{A}_{k+i}&\cdots&\mathbf{C}_{k+N}\\\end{bmatrix}.
\end{equation}

Furthermore, the cross-covariance matrices caused by the measurement in the time window can be defined as:
\begin{equation} \label{appendix_eq8}    \mathbcal{Q}_{l,k}\triangleq\mathbb{E}\left[\mathbcal{w}_l\mathbcal{w}_k^T\right],
\end{equation}
\vspace{-11pt}
\begin{equation} \label{appendix_eq9}
\mathbcal{R}_{l,k}\triangleq\mathbb{E}\left[\mathbcal{v}_l\mathbcal{v}_k^T\right].
\end{equation}

Finally, a matrix that contains all error covariance matrices involved in the smoothing process is defined below for the ease of implementation:
\begin{equation} \label{appendix_eq10}
    \mathbf{\Lambda}_k\triangleq
    \begin{bmatrix}\mathbf{P}_{k-1} & \mathbf{P}_{k-1}^{\mathbf{xw}} & \mathbf{P}_{k-1}^{\mathbf{xv}}\\
    \mathbf{P}_{k-1}^{\mathbf{wx}} & \mathbcal{Q}_{k,k} & \mathbf{P}_k^{\mathbf{wv}}\\
    \mathbf{P}_{k-1}^{\mathbf{vx}} & \mathbf{P}_k^{\mathbf{vw}} & \mathbcal{R}_{k,k}\\ \end{bmatrix}.
\end{equation}
\clearpage
\bibliography{ref} 





\end{document}